\documentstyle[aps,prd,amssymb]{revtex}
\input epsfig.sty
\def\laq{\raise 0.4ex\hbox{$<$}\kern -0.8em\lower 0.62 ex\hbox{$\sim$}}
\def\gaq{\raise 0.4ex\hbox{$>$}\kern -0.7em\lower 0.62 ex\hbox{$\sim$}}

\draft

\begin{document}

\title{Hypermagnetic Knots, Chern-Simons Waves and the Baryon Asymmetry}

\author{ Massimo Giovannini\footnote{
Electronic address: giovan@cosmos2.phy.tufts.edu}}
\address{{\it Department of Physics 
and Astronomy, Tufts University, Medford, Massachusetts 02155, USA}}

\maketitle
\begin{abstract}
At finite hyperconductivity and 
finite fermionic density the flux lines of long range hypermagnetic 
fields may not have a topologically trivial structure. 
The combined evolution of the chemical potentials and of pseudoscalar 
fields (like the axial Higgs), possibly present for temperatures in 
the TeV range, can twist the hypercharge 
flux lines, producing, ultimately,  hypermagnetic knots (HK). 
The dynamical features of the HK depend upon the various particle 
physics parameters of the model (pseudoscalar masses and couplings, 
strength of the electroweak phase transition, hyperconductivity  
of the plasma) and upon the magnitude of the primordial flux sitting 
in topologically trivial configurations of the hypermagnetic field.
We study different cosmological scenarios  where HK can be 
generated. We argue that the fermionic number sitting in HK 
can be released producing a seed for the 
Baryon Asymmetry of the Universe (BAU) provided the typical scale of the
knot is larger than the diffusivity length scale.
We derive constraints on  the primordial hypermagnetic flux
required by our mechanism and we provide a measure of the 
parity breaking by connecting the degree of knottedness of the 
flux lines to the BAU. We rule out the ordinary axion as a possible 
candidate for production (around temperatures of the order of the GeV) 
of {\em magnetic} knots since the produced 
{\em electromagnetic}  helicity is negligible  
(for cosmological standard) if the initial amplitude of the axion 
oscillations is of the order of the Peccei-Quinn breaking scale. 

\end{abstract}
\vskip0.5pc
\noindent

\renewcommand{\theequation}{1.\arabic{equation}}
\setcounter{equation}{0}
\section{Formulation of the Problem}

In a globally neutral plasma, the
 transversality of the magnetic fields implies, 
at high  conductivity, the conservation of 
{\em both} magnetic flux and magnetic helicity. 
The plasma element evolves always glued together with 
magnetic flux lines whose  
 number of associated knots and twists does not
change . The {\em energetical}
and the {\em topological} properties of the
 magnetic flux lines are {\em conserved} in the superconducting (or ideal) 
regime. Moreover, the linearity of the magnetohydrodynamical equations 
in the mean electric and magnetic fields implies
that there are no chances  that large scale magnetic 
fields could be {\em generated} only using electromagnetic plasma effects. 

As noticed long ago \cite{fermi1,fermi2} magnetohydrodynamics (MHD) represents 
a quite powerful method in order to analyze the dynamics of the interstellar
plasma. In particular 
the topological properties of the magnetic field lines and of the bulk
velocity field play a significant role in the dynamo theory 
(often invoked in order to explain the typical amplitudes of large 
scale magnetic fields in spiral galaxies \cite{rev1}) and 
 it seems  well established that for 
successfully implementing  the dynamo mechanism we 
are led to postulate a {\em global} parity violation at the 
scale at which the dynamo action is turned on \cite{parker,rev3}. 
Again, 
because of the linearity of MHD, in order to explain the galactic
 magnetic fields 
\cite{rev1,rev3} we have to assume that at some finite 
moment of time (after the Big-Bang) some magnetic seeds \cite{seed} had to be 
present.

Up to now, the attention  has been mainly focused
on the energetical properties of the seeds  and very little has been said 
concerning their topological features. Apart 
from the dynamo mechanism, the topology of the magnetic flux lines 
seems less relevant (for the seed problem) than their associated energy 
density.  However, going backward in time, we find that the topological 
properties of long range (Abelian) gauge fields can have unexpected 
physical implications.
 
The importance of the 
 topological properties of long range (Abelian) hypercharge magnetic fields 
has been indeed  emphasized in the past \cite{vil,hyp}. In \cite{vil} the 
interesting possibility of macroscopic parity breaking at finite fermionic 
density and finite temperature was studied and it was
suggested that if parity is globally broken an electromagnetic 
current density directed along the magnetic field can arise.
A similar observation has been made in the context of
the standard electroweak theory at finite temperature and chemical potential
 \cite{hyp} where it has been observed
 that, thanks to the anomalous coupling of the hypercharge 
 the fermionic density can be converted 
into infra-red modes of the hypercharge field.
 
If the spectrum of hypermagnetic 
fields is dominated by parity non-invariant configurations
 the baryon asymmetry of the Universe could be the result of 
the decay of these condensates \cite{m1}.  The purpose of the present
paper is to connect the topological
properties of the hypermagnetic flux tubes to the   
Baryon Asymmetry of the  Universe (BAU) by discussing a class of dynamical
mechanisms leading to the generation of HK.

The evolution of mean hypermagnetic fields in an electroweak plasma at 
finite conductivity and finite fermionic densities can be  derived \cite{m2} 
and it turns out that a good gauge-invariant measure of global parity 
invariance is  the hypermagnetic helicity, i.e. 
$\vec{{\cal H}}_{Y} \cdot\vec{\nabla} \times\vec{\cal H}_{Y} $. 
If the hypermagnetic field distribution is {\em topologically 
trivial} (i.e.
$\langle\vec{\cal H}_{Y} \cdot\vec{\nabla} \times\vec{\cal H}_{Y}\rangle =0$) 
matter--antimatter fluctuations can be generated at the electroweak scale 
\cite{m3}. Provided the
 typical correlation scale of the magnetic helicity is larger
than (or of the order of) the neutron diffusion scale blue-shifted at the 
electroweak epoch [i.e. $L_{n}(T_{ew}) \sim 0.3 ~{\rm cm}$ 
at $T_{ew} \sim 100~{\rm GeV}$],
 matter--antimatter domains can survive until the onset 
of Big-Bang nucleosynthesis (BBN) leading to non standard initial conditions 
for the light element abundances calculations. In the presence 
of a fraction of antimatter at BBN a reduction of the neutron
to proton ratio can be foreseen with consequent reduction of the $^{4}$He 
abundance. This effect was recently pointed out in BBN calculations 
\cite{RJ} in the presence of spherical antimatter domains and further work
 is in progress \cite{HS}.

An example of {\em topologically trivial}  distribution of 
hypermagnetic fields leading to some of the above mentioned effects
 is a stochastic bacground
whose (parity-invariant) two-point function can be expressed, 
in Fourier space, as
\begin{equation}
G_{ij}(k)= 
k^2f(k)\delta^{(3)}(\vec{k} -\vec{k}') 
\biggl(\delta_{ij} - \frac{{k_i} k_{j}}{k^2}\biggr), ~~~f(k) = 
\frac{{\cal A}(k)}{k}
\label{stoc}
\end{equation}
(${\cal A}(k)$ 
is a dimension-less factor measuring the intensity of the stochastic
 hypermagnetic background).
Since hypermagnetic fields are strictly divergence free (i.e. no 
hypermagnetic monopoles are present) the magnetic flux lines form a mixture 
of closed loops evolving glued to their electroweak plasma element traveling 
with velocity $\vec{v}$. As far as we are
 concerned with scales larger than the hypermagnetic diffusivity scale 
(i.e. $L_{\sigma} \sim 3 \times 10^{-9} {\rm cm}$ at $T_{{\rm ew}}$), then 
these flux loops cannot break or twist.

If the Universe is filled, at some scale, by {\em topologically non-trivial}
HK  the flux  lines form a network of {\em stable knots 
and twists} which cannot be broken provided the conductivity is high.
Therefore $\langle\vec{\cal H}_{Y} \cdot\vec{\nabla} \times\vec{\cal H}_{Y}
\rangle\neq 0 $. Thus, 
completely homogeneous
hypermagnetic configurations cannot be topologically non-trivial.
If we want to have knots and twists in the flux lines, we 
are necessarily driven towards field configurations which are not only 
time but also space dependent like the Chern-Simons waves 
\cite{hyp,m1} analogous to the magnetic knot configurations \cite{m2} 
frequently employed in the analysis of dynamo instabilities \cite{rev3,ma}.

If the hypermagnetic configurations are {\em topologically non-trivial} (and 
consequently inhomogeneous) some knots in the lines of the velocity  field
can be also expected. Indeed, by solving the 
hypermagnetic diffusivity equation  to leading order in the resistivity
expansion, the velocity field turns out to be proportional to the 
hypermagnetic field, namely $\vec{v}_{Y}\simeq 
\vec{\cal H}_{Y}/(\sqrt{N_{eff}} T_{ew}^2)$.
Notice that this is nothing but  the electroweak analogous of the 
Alfv\'en velocity \cite{bis} so important in the spectrum of (ordinary) plasma 
waves \cite{mhd}.  
Since the hypermagnetic fields are divergence-free, the fluid will also be 
incompressible (i.e. $\vec{\nabla}\cdot\vec{v}_{Y} =0$) as required, 
by equations of MHD. It is now clear that the topological properties of
the magnetic flux lines can be connected with the topological 
properties of the velocity field lines. 

In this paper we want to understand if it is at all possible to {\em generate
a topologically non-trivial distribution of hypercharge fluctuations 
from a topologically trivial collection of the same fields}.
More physically we are going to investigate how to create knots and twists 
in a stochastic collection of disconnected magnetic flux loops. 
We would also like to understand if this kind of mechanism can or cannot
be implemented in some reasonably motivated cosmological framework. 

A way of twisting the topology of a collection of independent 
(hypermagnetic) flux loops,
is to have dynamical pseudoscalar fields at finite fermionic density. 
The reason is twofold.
On one hand dynamical pseudoscalars fields are coupled to 
gauge  fields through the anomaly and, as a result, they induce a coupling 
between  the two polarizations of photons and/or hyperphotons. 
On the other hand, 
unlike the ordinary electromagnetic fields, the 
coupling of the hypercharge field to fermions is 
chiral \cite{hyp}. The
generation of hypermagnetic knots can influence the evolution 
of the chemical potential leading  
 to the production of the BAU \cite{m1}.
The BAU can be produced 
{\em provided} the electroweak phase transition (EWPT) is
strongly first order, {\em provided} the typical
scale of the HK exceeds the diffusivity scale associated 
with the finite value of the hyperconductivity, and 
{\em provided} the rate of the slowest processes in the plasma is 
sufficiently high. Unfortunately, in the minimal standard model
(MSM) the EWPT cannot be strongly  first order for 
Higgs boson masses larger than the $W$ boson mass \cite{PT1}.
It is true that the 
presence of a (topologically trivial) hypermagnetic 
background \cite{PT0,PT2} can modify the phase 
diagram but cannot make the EWPT strongly first order. We will
argue that our scenario can have some chances in the context of the 
minimal supersymmetric standard model (MSSM) where, incidentally,
the rate of the right-electron chirality flip processes can also
be larger  than in the MSM.

Dynamical
pseudoscalars have been used in the past in order to generate 
seeds for the magnetohydrodynamical evolution
\cite{jak,carr}. Our concern here is different and it is mainly 
connected with  the BAU. Suppose that
at some moment in the life of the Universe dynamical pseudoscalars
fields are present. Then,
at finite conductivity and finite fermionic density 
(i.e. finite chemical potential) both,
 the chemical potential and the pseudoscalar particles, can be 
coupled to the anomaly. 

In flat space, the interactions between a 
pseudoscalar field and an  Abelian gauge field can be parameterized as 
\begin{equation}
c\frac{\psi}{4 M} {\cal Y}_{\alpha\beta} {\tilde{{\cal Y}}}^{\alpha\beta}
\label{psint}
\end{equation}
where $c$ is the pseudoscalar coupling constant, $M$ the typical energy 
scale and 
${\cal Y}_{\alpha\beta} =\partial_{[\alpha} {\cal Y}_{\beta]}$ is the gauge 
field strength expressed in terms of the vector potentials. In our 
analysis we will be mainly concerned with the case of the hypercharge 
field and with the case of the ordinary electromagnetic field.
Imagine now that the pseudoscalar field is uniform but time dependent.  
If our initial state is a (topologically trivial) 
stochastic distribution of gauge field  fluctuations the 
effect of the interaction described in Eq. (\ref{psint}) will be 
to couple the two (transverse) polarizations of the 
gauge fields. Since  the evolution 
equations of the two (independent) polarizations are 
different,  there are no reasons to expect that 
the topological structure of the initial gauge 
field distribution will be left unchanged by the 
presence of pseudoscalar interactions. 
The inclusion of a (macroscopic) ohmic conductivity 
will automatically select a preferred reference 
frame (the so called ``plasma frame'') where the 
flux and helicity associated with the magnetic 
component of the the Abelian background are both 
(approximately) conserved if the resistivity
is small. 

In the plasma frame, the initial 
stochastic distribution of gauge field fluctuations
can be viewed as a collection of unknotted 
magnetic flux tubes (closed by transversality)
which evolve independently without
breaking or intersecting each others. The inclusion 
of pseudoscalar interactions will then alter 
the topological properties of the magnetic flux 
lines by introducing links between independent 
loops and by also twisting a single loop.
Part of the results reported in this paper have been recently presented 
(in a more compact form) in \cite{mmm}.

The 
plan of our paper is then the following. In Section II we introduce 
the basic description of hypermagnetic fields coupled to dynamical
 pseudoscalars
in an empty (but expanding) space. 
We will then solve the time evolution of the field
operators in the Heisenberg representation. 
In Section III we will formally address the problem 
of emission of  hypermagnetic knots induced by 
dynamical pseudoscalars.
In Section IV we will concentrate our attention on the 
hypermagnetic helicity production during an inflationary phase 
whereas in Section V we will discuss the radiation dominated phase.
In Section VI we will connect the production of (mean) hypermagnetic helicity
to the  BAU and  in Section VII 
we will investigate the regions of parameter space where the generate BAU 
can be phenomenologically relevant.
In Section VIII we will  examine, for sake of completeness, 
the problem of global parity breaking induced by axionic particles. 
Section IX contains our concluding remarks. In the Appendix we collected some 
useful technical results.

\renewcommand{\theequation}{2.\arabic{equation}}
\setcounter{equation}{0}
\section{Hypermagnetic fields and Chern-Simons waves in empty space}

The Abelian nature of the hypercharge field does not imply that 
the hypermagnetic flux lines should have a trivial topological 
structure.
We will name {\em topologically trivial} the configurations 
whose field lines are {\em closed} (by transversality of the field) 
and {\em unknotted}. Conversely, in
 the gauge ${\cal Y}_0=0$, $\vec{\nabla}\cdot\vec{{\cal Y}}=0$, 
an example of 
 topologically non-trivial configuration of the hypercharge field is 
the Chern-Simons wave \cite{hyp,m1,m2}
\begin{eqnarray}
{\cal Y}_{x} (z, t) &=& {\cal Y}(t) \sin{k_0 z},
\nonumber\\
{\cal Y}_{y} (z, t) &=& {\cal Y}(t) \cos{k_0 z}, 
\nonumber\\
{\cal Y}_{x} (z,t) &=& 0.
\label{conf1}
\end{eqnarray}
This particular configuration is not homogeneous but it
 describes a hypermagnetic knot with {\em homogeneous} helicity and 
 Chern-Simons number  density
\begin{eqnarray}
&&\vec{{\cal H}}_{Y} \cdot \vec{\nabla} \times\vec{\cal H}_{Y} = 
k_0 {\cal H}^2(t),
\nonumber\\
&&n_{CS} = - \frac{g'^2}{32 \pi^2} \vec{{\cal H}}_{Y}\cdot \vec{{\cal Y}} = 
\frac{ g'^2 }{32 \pi^2 k_0} {\cal H}^2(t),
\end{eqnarray}
where $\vec{{\cal H}}_{Y} 
=\vec{\nabla} \times \vec{{\cal Y}}$, ${\cal H}(t) = k_0 
{\cal Y}(t)$; $g'$ is the $U(1)_{Y}$ coupling. Other examples can be found. 
For instance, 
it is possible to construct hypermagnetic knot configurations 
with finite energy and helicity which are localized in space within a  
typical size $L_{s}$. These examples are reported in Appendix A.

Suppose now that the Universe can be described, at some epoch below 
the Planck scale, 
by a homogeneous, isotropic and spatially flat 
Friedmann-Robertson-Walker (FRW) metric with line element
\begin{equation}
ds^2 = g_{\mu\nu} dx^{\mu} dx^{\nu},~~~g_{\mu\nu} = a^2(\tau)\eta_{\mu\nu}
\label{metric}
\end{equation}
($\tau$ is the conformal time coordinate and $\eta_{\mu\nu}$ is the
ordinary flat metric with signature $[+, -, -, -]$). 
Let us assume that  dynamical pseudoscalar
particles are  evolving in the background geometry given by Eq. (\ref{metric}). 
The pseudoscalars are {\em not} a source of the background (i.e. they do
not affect the time evolution of the scale factor) but, nonetheless,
they evolve according to their specific dynamics and can excite 
other degrees of freedom.

The effective action describing the interaction of a dynamical 
pseudoscalars with hypercharge fields can be written,
in curved space as
\begin{equation}
S= \int d^{4} x \sqrt{-g} \biggl[ \frac{1}{2}g^{\alpha\beta}
\partial_{\alpha}\psi \partial_{\beta}\psi  - V(\psi) -
\frac{1}{4}Y_{\alpha\beta}Y^{\alpha\beta} + 
c\frac{\psi}{4 M}
Y_{\alpha\beta}\tilde{Y}^{\alpha\beta}\biggr].
\label{action}
\end{equation}
This action is quite generic. In the case $V(\psi) = (m^2/2) \psi^2$
Eq. (\ref{action}) is  nothing but the curved 
space generalization of the model usually employed in direct searches
of axionic particles \cite{sikivie}. The constant in front of the anomaly 
is a model-dependent factor. For example, in the case of axionic particles
, for large temperatures $T \geq m_{W}$,
 the Abelian gauge fields present in Eq. (\ref{action}) will be 
hypercharge fields and $c = c_{\psi Y} \alpha'/(2 \pi)$ where $\alpha'=
g'^2/4\pi$ and $c_{\psi Y}$ is a numerical factor of order $1$ 
which can be computed (in a specific axion scenario) by knowing the
 Peccei-Quinn charges of all the fermions present in the model \cite{kim}. 
For small temperatures 
 $T\leq m_{W}$ we have that the Abelian fields present in the 
action (\ref{action}) will coincide with ordinary  electromagnetic 
fields and
$c = c_{\psi\gamma} \alpha_{{\rm em}}/2\pi$ where $\alpha_{{\rm em}}$ 
is the fine structure constant and $c_{\psi\gamma}$ 
is again a numerical factor.

The coupled system of equations describing the evolution of the 
pseudoscalars 
and of the Abelian gauge fields can be easily derived by varying the action 
with respect to $\psi$ and $Y_{\mu}$,
\begin{eqnarray}
&&\Box\psi + \frac{\partial V}{\partial\psi}= \frac{c}{4 M} Y_{\alpha\beta} 
\tilde{Y}^{\alpha\beta},
\nonumber\\
&&\nabla_{\mu}  Y^{\mu\nu} = 
\frac{c}{M}\nabla_{\mu}\psi\tilde{ Y}^{\mu\nu},~~~
\nabla_{\mu} \tilde{ Y}^{\mu\nu}=0,
\label{eqmotion}
\end{eqnarray}
where,
\begin{equation}
\nabla_{\mu} Y^{\mu\nu} = \frac{1}{\sqrt{-g}}
 \partial_{\mu}\biggl[ \sqrt{-g} Y^{\mu\nu}\biggr],~~~\nabla_{\mu}
 \tilde{Y}^{\mu\nu} = \frac{1}{\sqrt{-g}}\partial_{\mu} \biggr[
 \sqrt{-g} \tilde{Y}^{\mu\nu}\biggr],
\end{equation}
are the usual covariant derivatives defined from the background metric
of Eq. (\ref{metric}) and 
$\Box\psi= \nabla_{\alpha}\nabla^{\alpha}\psi =[-g]^{-1/2}
 \partial_{\alpha}[\sqrt{-g}
g^{\alpha\beta}\partial_{\beta} \psi]$.
Recall now that
\begin{equation}
Y_{0i}= a^2 {\cal E}_{i},~~~\tilde{Y}_{0i} = a^2 {\cal B}_{i},~~~ 
Y_{i j}= -a^2 \epsilon_{ijk} {\cal B}_{k},~~~
\tilde{Y}_{ij} = a^2 \epsilon_{ijk}  {\cal E}_{k}
\label{def}
\end{equation}
where ${\cal E}_{i}$ and ${\cal B}_{i}$ are the flat space fields
(the contravariant components obtained by raising
the indices with the metric given in Eq. (\ref{metric})). Using
 Eq. (\ref{def}), 
Eqs. (\ref{eqmotion}) can be written in terms of the physical gauge  fields
\begin{eqnarray}
&&\psi'' + 2 {\cal H} \psi' - \nabla^2\psi 
+ a^2 \frac{\partial V}{\partial \psi} 
= - \frac{1}{a^2}\frac{c}{M} \vec{E}_{Y}\cdot \vec{B}_{Y}, ~~~
{\cal H}= \frac{a'}{a},
\nonumber\\
&&\vec{\nabla}\cdot\vec{B}_{Y} =0,
~~~\vec{\nabla}\times\vec{E}_{Y} +\vec{B}_{Y}' =0,~~~
\vec{\nabla}\cdot\vec{E}_{Y} = \frac{c}{M}\vec{\nabla}\psi \cdot\vec{B}_{Y},
\nonumber\\
&&\vec{\nabla}\times\vec{B}_{Y} = \vec{E}_{Y}' 
- \frac{c}{M} \biggl[ \psi' \vec{B}_{Y} 
+ \vec{\nabla}\psi\times \vec{E}_{Y}\biggr],
\label{system}
\end{eqnarray}
(notice that the prime denotes differentiation with respect to the 
conformal time ).
The rescaled fields  $\vec{E}_{Y}$ and $\vec{B}_{Y}$ are related to
$\vec{{\cal E}}_{Y}$ and $\vec{{\cal H}}_{Y}$ as $\vec{E}_{Y} 
= a^2 \vec{{\cal E}}_{Y}$,
$\vec{B}_{Y} = a^2 \vec{{\cal B}}_{Y}$. 

We want now to study the amplification of gauge field 
fluctuations induced by 
the time evolution of $\psi$. Then, the evolution equation for the 
hypermagnetic fluctuations $\vec{H}_{Y}$ can be obtained by
 linearizing  Eqs. (\ref{system}). We will assume that any background 
gauge field is absent.
 In the linearisation procedure we will also assume that the pseudoscalar 
field can be treated as completely homogeneous 
(i.e.$ |\vec{\nabla}\psi|\ll \psi'$). This seems to be natural if, 
prior to the radiation dominated epoch, an inflationary phase diluted 
the gradients of the pseudoscalar.

In this approximation, the result of the linearization can be simply written
in terms of the vector potentials in the gauge  
$Y^0 =0$ and $\vec{\nabla}\cdot\vec{Y} =0$:
\begin{eqnarray}
&&\vec{Y}'' - \nabla^2  \vec{Y} +\frac{c}{M} \psi' \vec{\nabla} \times 
\vec{Y} =0,
\label{vectorp}\\
&& \ddot{\psi} + 3 H\dot{\psi} + \frac{\partial V}{\partial\psi} =0, ~~~
H=\frac{\dot{a}}{a}\equiv \frac{{\cal H}}{a},
\label{psi}    
\end{eqnarray}
where the over-dot denotes differentiation with respect to the cosmic 
time coordinate $t$ [i.e. $a(\tau) d\tau = d t$]\footnote{ In order to avoid 
confusions we denoted with $H$ the Hubble factor is cosmic time and with 
${\cal H}$ the Hubble factor in conformal time. These quantities 
cannot be confused with $\vec{H}_{Y} = a^2 \vec{\cal H}_{Y}$ since, when 
some ambiguity might arise, we will always keep the appropriate 
subscripts.}.
By combining the evolution equations for the gauge fields we can  find a
decoupled evolution equation for $\vec{H}_{Y}= \vec{\nabla} \times \vec{Y}$,
\begin{equation}
\vec{H}_{Y}'' - \nabla^2\vec{H}_{Y} + \frac{c}{M} \psi' \vec{\nabla}
 \times\vec{H}_{Y} =0,
\end{equation}
and we can  also decompose the vector potentials in Fourier integrals
\begin{equation}
\vec{ Y}(\vec{x},\tau) = \sum_{\alpha}\int \frac{d^3 k}{\sqrt{(2\pi)^3}} 
\Biggl(Y_{\alpha}(k,\tau) \vec{e}_{\alpha} e^{i\vec{k} \cdot \vec{x}} +
 Y^{\ast}_{\alpha}(k,\tau) \vec{e}_{\alpha}  e^{-i\vec{k} \cdot
\vec{x}}\Biggr),
\label{decomposition1}
\end{equation}
where $\alpha=1,2$ and runs over the two (real) linear polarization vectors
$\vec{e}_{\alpha}$ whose direction depends on the propagations direction 
$\vec{k}$. According to this decomposition,  $\vec{e}_{1}$, $\vec{e}_{2}$ and
 $\vec{e}_3=\vec{k}/|\vec{k}|$ form a set of three mutually
orthonormal unit vectors (i.e. $\vec{e}_{1} \times \vec{e}_{2} =
 \vec{e}_{3}$).
Inserting Eq. (\ref{decomposition1}) into Eq. (\ref{vectorp}) we get a
 system of (still coupled) equations for the Fourier components
\begin{equation}
Y_{1}'' + k^2 Y_{1} - i k \frac{\psi'}{M} Y_{2} =0
,~~~~Y_{2}'' + k^2 Y_{2} + i k \frac{\psi'}{M} Y_{1} =0,
\label{sys2}
\end{equation}
In order to decouple this system of equations we can pass from {\em linear 
polarization} vectors to (complex) {\em circular plarizations} by defining:
\begin{equation}
Y_{\pm}(k,\tau) = Y_1(k,\tau) \pm i Y_2(k,\tau),~~~
Y^{\ast}_{\pm}(k,\tau) = Y^{\ast}_1(k,\tau) \mp i Y^{\ast}_2(k,\tau).
\end{equation}
In terms of these linear combinations the decomposition of 
(\ref{decomposition1}) becomes
\begin{equation}
\vec{{Y}}(\vec{x},\tau) = \frac{1}{\sqrt{2}}\sum_{\beta}\int\frac{d^3 k} 
{\sqrt{(2\pi)^3}}
\Biggl({ Y}_{\beta}(k,\tau) \vec{\epsilon}_{\beta}e^{i \vec{k}\cdot\vec{x}} 
+ {Y}^{\ast}_{\beta}(k,\tau)
\vec{\epsilon}^{~~\ast}_{\beta} e^{-i \vec{k}\cdot\vec{x}}\Biggr).
\label{decomposition2}
\end{equation}
where now $\beta=+,-$ labels the two circular polarizations
\begin{equation}
\vec{\epsilon}_{+} = \frac{\vec{e}_1 - i \vec{e}_{2}}{\sqrt{2}}, ~~~
\vec{\epsilon}_{-} = \frac{\vec{e}_1 + i \vec{e}_{2}}{\sqrt{2}},
\label{circular}
\end{equation}
satisfying the usual identities
\begin{eqnarray}
&&\vec{\epsilon}_{+} \cdot \vec{\epsilon}_{-}=1,
~~~\vec{\epsilon}^{~~\ast}_{+} = \vec{\epsilon}_{-},~~~
\vec{\epsilon}^{~~\ast}_{-} = \vec{\epsilon}_{+},
\nonumber\\
&&\vec{\epsilon}_{+} \cdot\vec{\epsilon}_{+} =0,~~~
\vec{\epsilon}_{-} \cdot\vec{\epsilon}_{-} =0,
\nonumber\\
&&\vec{k}\times \vec{\epsilon}_{+} = i |\vec{k}| \vec{\epsilon}_{+},~~~
\vec{k}\times \vec{\epsilon}_{-} = -i |\vec{k}| \vec{\epsilon}_{-}.
\label{relations}
\end{eqnarray}
In terms of $Y_{\pm}$ Eqs. (\ref{sys2}) can be written as 
\begin{equation}
{Y}_{\pm}'' + \omega_{\pm}^2 {Y}_{\pm} =0,
~~~{{Y}^{\ast}}_{\pm}'' + \omega_{\pm}^2 {Y}^{\ast}_{\pm} =0,~~~
\omega^2_{\pm} = k^2 \mp k \frac{c}{M} \psi'.
\label{decoupled}
\end{equation}
Notice that $Y_{\pm}$ are  the normal modes of the action 
describing the 
(classical) hypercharge fluctuations excited by the evolution of 
the dynamical
pseusoscalar. In other words the action (\ref{action}) is not diagonal in terms
of $Y_{1,2}(\vec{x}, \tau)$ but it can be diagonalised in terms of $Y_{\pm}$.
  By inserting the decomposition given 
in Eq. (\ref{decomposition2}) back  into the action (\ref{action}),
 by using 
Eqs. (\ref{relations}) and by 
integrating over $d^{3} x$ we obtain that the action for the hypercharge 
fluctuations can be expressed as:
 \begin{equation}
S= \int d \tau L(\tau),~~~ L(\tau)= \int d^3 k {\cal L}(k,\tau),
\end{equation}
with
\begin{equation}
{\cal L}(k,\tau) =  Y'_{+}(k,\tau) Y^{\ast\prime}_{+}
(k,\tau) + Y'_{-}(k,\tau) Y^{\ast\prime}_{-}(k,\tau) 
 -\omega^2_{+}Y_{+}(k,\tau) Y^{\ast}_{+}(k,\tau) 
-\omega^2_{-}Y_{-}(k,\tau)Y^{\ast}_{-}(k,\tau)
\label{diagonalaction}
\end{equation}
[notice that $Y^{\ast}_{+}(k,\tau)= Y_{+}(-k,\tau),~~~
Y^{\ast}_{-}(k,\tau)=Y_{-}(-k,\tau)$].

In order to make more explicit the meaning  of Eqs. (\ref{vectorp}) and 
(\ref{psi}) let us start from some qualitative considerations.
Suppose  that the 
potential term appearing in the action (\ref{action}) is exactly 
$(m^2/2)\psi^2$. Moreover, for sake of simplicity we will deal 
with the case where the conductivity is completely 
absent.  If $H\gg m$ the $\psi$ field freely evolves, and,
 as a consequence, we will have, from Eqs. 
(\ref{system})-(\ref{decoupled}) 
\begin{equation}
\psi' \sim \frac{1}{a^2},~~~
Y_{\pm}'' + \biggl[ k^2 \mp  k \frac{c}{a^2 M}\biggr] Y_{\pm} =0,  
\end{equation}
and similarly for the other complex conjugate solutions.
At $H\sim m$, $\psi$ will start oscillating with typical frequency $m$ and 
typical amplitude $\psi_0$:
\begin{equation}
\psi(t) \simeq \frac{\psi_0}{a^{3/2}}\sin{[m (t - t_0)]}.
\end{equation}
Since $H\leq m$ the effect of the Universe expansion can be neglected during
 the oscillating phase as long as we limit ourselves to time intervals 
$\Delta t \laq H^{-1}$. Moreover, since $\psi' = a \dot{\psi}$ from
 Eq. (\ref{decoupled}) we can see that for $\dot{\psi} >\omega c/M $, 
$Y_{+}$ will be amplified
(we defined $\omega = k/a$ as the physical frequency).
The maximal amplified frequency will be $\omega_{{\rm max}} \simeq
\dot{\psi}_{{\rm max}} c/M \sim c m(\psi_0/M)$. In the same regime $Y_{-}$
 is not amplified.
 Then it is not unreasonable to expect that pseudoscalar quantities 
(like $\vec{Y}\cdot\vec{\nabla}\times\vec{Y}$, 
$\vec{H}_{Y}\cdot\vec{\nabla}\times\vec{H}_{Y}$ etc.) can be produced.
Of course, the fact that pseudoscalar configurations of the hypercharge fields 
are produced does not  necessarily mean that they cannot be erased by the 
finite value of the conductivity. We will come back on this point later.

\renewcommand{\theequation}{3.\arabic{equation}}
\setcounter{equation}{0}
\section{Hypermagnetic Helicity Amplification}

Starting from a stochastic mixture 
of topologically trivial fluctuations, the time evolution of a generic 
dynamical pseudoscalar excites gauge field configurations which are
topologically non-trivial. 
Suppose that the gauge field fluctuations are in their vacuum state, 
and suppose that after a period of free-rolling the pseudoscalar field 
starts oscillating at a curvature scale $H\sim m$. This can happen both in
towards the end of an inflationary epoch and/or in a radiation dominated 
regime, as we will show in the next two Sections. 

Particle creation in an external field \cite{birrel} 
plays an important role in the analysis of graviton production due to the
evolution of the gravitational background \cite{gris} and it is also
crucial for a rigorous treatment of the normalization of density
fluctuations in the context of ordinary inflationary models \cite{rev4}. In
quantum optics \cite{qo} similar techniques are used in order to 
describe the photon production in an external (pump) laser
 field interacting with a non-linear material. In condensed matter theory
 these techniques are also employed and their similarities with 
cosmological problems are certainly worth to emphasize \cite{vol}.  

In the Heisenberg 
description the process of amplification can be understood in terms of
unitary transformations (the  Bogoliubov-Valatin transformations)
 relating the field operators between two asymptotic vacua.
In the  Schroedinger  picture the amplification process is described
through the action of a time evolution operator connecting the initial 
vacuum state to the final (multiparticle) ``squeezed''  vacuum state 
\cite{gris,qo,mg}.

The classical (canonical) Hamiltonian for the gauge field fluctuations 
can be 
obtained from the canonical action of Eq. (\ref{diagonalaction}) 
\begin{equation}
H(\tau) = \int d^{3} k {\cal H}(k,\tau),~~~ {\cal H}(k,\tau)=
\pi_{+}^{\ast}\pi_{+} + \pi^{\ast}_{-}\pi_{-} 
+ \omega^2_{+}Y^{\ast}_{+}Y_{+} + \omega^2_{-}Y^{\ast}_{-}Y_{-},
\label{ham}
\end{equation}
where $\pi_{\pm}$ and $\pi^{\ast}_{\pm}$ are the canonical momenta and where 
$\omega_{\pm}^2$ are not necessarily positive definite.
 From Eq. (\ref{ham}) the 
evolution equations (\ref{decoupled}) can be directly obtained in the 
Hamiltonian formalism. 
Our problem is analogous to the quantization
of two decoupled (complex) harmonic oscillators and it is closely
analogous to the quantization of the two circular polarization of the
electromagnetic field in the radiation gauge. The only (but crucial)
difference is that  the two polarizations obey (in our
case) two different evolution equations (i.e. $\omega^2_{+}\neq
\omega^2_{-}$). In the limit $\psi'=0$ (i.e. $\omega_{+} =\omega_{-}$)
the ordinary ``electromagnetic'' case is completely recovered \cite{qo}.

Promoting the classical normal modes to quantum mechanical operators
\begin{eqnarray}
\pi_{+}(k,\tau) \rightarrow
\hat{\pi}_{+}(k,\tau),~~~Y_{+}(k,\tau)\rightarrow \hat{Y}_{+}(k,\tau),
\nonumber\\
\pi_{-}(k,\tau) \rightarrow
\hat{\pi}_{-}(k,\tau),~~~Y_{-}(k,\tau)\rightarrow 
\hat{Y}_{-}(k,\tau),
\end{eqnarray}
we can impose the commutation relations
\begin{equation}
\bigl[\hat{Y}_{\beta}(k,\tau) ,\hat{\pi}_{\beta'}(p,\tau)\bigr] = 
i \delta_{\beta\beta'}\delta^{(3)}(\vec{k}- \vec{p}),~~~
\bigl[\hat{Y}_{\beta}(k,\tau) ,\hat{Y}_{\beta'}(p,\tau)\bigr]=0,~~~
[\hat{\pi}_{\beta}(k,\tau) ,\hat{\pi}_{\beta'}(p,\tau)] =0
\end{equation}
where $\beta=+,-$.
The dynamical evolution follows the Heisenberg
equations of motion given by
\begin{equation}
\hat{Y}'_{\beta}=  i\bigr[ \hat{H}, \hat{Y}_{\beta}\bigl],~~~
\hat{\pi}'_{\beta}=  i\bigl[ \hat{H}, \hat{\pi}_{\beta}\bigr],
\end{equation}
where $\hat{H}$ is the Hamiltonian operator.
A consequence of these equations is that the operator $\hat{Y}_{\pm}$
obey, in the Heisenberg representation, the same evolution equation
of the related classical quantity derived in  Eqs. (\ref{decoupled}). 

We define the creation and annihilation operators as
\begin{eqnarray}
&&\hat{Y}_{+}(\vec{k},\tau)= \frac{1}{\sqrt{2 \omega_{+}}}\biggl[ 
\hat{a}_{+}(-\vec{k},\tau) + \hat{a}^{\dagger}_{+}(\vec{k},\tau)\biggr],
~~~\hat{Y}_{-}(\vec{k},\tau)= \frac{1}{\sqrt{2 \omega_{-}}}\biggl[ 
\hat{a}_{-}(-\vec{k},\tau) + \hat{a}^{\dagger}_{-}(\vec{k},\tau)\biggr],
\nonumber\\
&&\hat{\pi}_{+}(\vec{k},\tau) =- i\sqrt{\frac{\omega_{+}}{2}}\biggl[
 \hat{a}_{+}(\vec{k},\tau)- \hat{a}^{\dagger}_{+}(-\vec{k},\tau)\biggr],
~~~\hat{\pi}_{-}(\vec{k},\tau) = -i\sqrt{\frac{\omega_{-}}{2}}\biggl[
 \hat{a}_{-}(\vec{k},\tau)- \hat{a}^{\dagger}_{-}(-\vec{k},\tau)\biggr].
\end{eqnarray}
Notice that $\hat{a}_{\pm}(-\vec{k},\eta)$ are annihilation operators
for an excitation with momentum $-\vec{k}$ whereas 
$\hat{a}_{\pm}(\vec{k},\eta)$ are annihilation operators
for an excitation with momentum $+\vec{k}$. The simultaneous presence
of both sets of operators guarantees the explicit three momentum
conservation of the whole formalism. The commutation relations obeyed
by these operators are the standard ones, namely
\begin{equation}
 \bigl[\hat{a}_{\beta}(\vec{k},\tau), 
\hat{a}^{\dagger}_{\beta'}(\vec{p},\tau)\bigr] = 
\delta_{\beta\beta'}\delta^{(3)}(\vec{k} - \vec{p}),~~~
 \bigl[\hat{a}_{+}(\vec{k},\tau), 
\hat{a}_{-}(\vec{p},\tau)\bigr] =0,
\end{equation}
and the Hamiltonian operator can then be written as 
\begin{eqnarray}
\hat{H}(\tau)= \sum_{\beta = -,+}\int d^3 k 
\Biggl[ \omega_{\beta}(\tau)\biggl( 
\hat{a}_{\beta}^{\dagger}(\vec{k},\tau)\hat{a}_{\beta}(\vec{k},\tau)
+\hat{a}_{\beta}^{\dagger}(-\vec{k},\tau)\hat{a}_{\beta}(-\vec{k},\tau)+
 1\biggr)\Biggr].
\end{eqnarray}
The field  operators  before the amplification took place can then
 be written as:
\begin{equation}
\hat{Y}_{+,in} = \biggl[\hat{a}_{+,in} f_{in} +
\hat{a}^{\dagger}_{+,in} f^{\ast}_{in} \biggr],
~~~\hat{Y}_{-,in} = \biggl[\hat{a}_{-,in} F_{in} +
\hat{a}^{\dagger}_{-,in} F^{\ast}_{in} \biggr].
\end{equation}
The same decomposition can be done for the operators after the
amplification processes has taken place:
\begin{equation}
\hat{Y}_{+,out} = \biggl[\hat{a}_{+,out} g_{out} +
\hat{a}^{\dagger}_{+,out} g^{\ast}_{out} \biggr],~~~
\hat{Y}_{-,out} = \biggl[\hat{a}_{-,out} G_{out} +
\hat{a}^{\dagger}_{-,out} G^{\ast}_{out} \biggr].
\end{equation}
We can formally write the out-going 
mode functions in terms of the
in-going mode functions :
\begin{eqnarray}
&&g_{out}(\tau) = c_{+} f_{in}(\tau) + c_{-} f^{\ast}_{in}(\tau),
\nonumber\\
&&G_{out}(\eta) = \tilde{c}_{+} F_{in}(\tau) + \tilde{c}_{-}
F^{\ast}_{in}(\tau).
\label{transf}
\end{eqnarray}
Since $\hat{Y}^{\pm,in}$ and $\hat{Y}^{\pm, out}$ represent 
the same solution they must be equal:
\begin{equation}
\hat{Y}_{+,in} = \hat{Y}_{+,out},~~~\hat{Y}_{-,in} = \hat{Y}_{-,out}.
\end{equation}
Inserting the explicit form of the solutions we have that
\begin{eqnarray}
&&\hat{a}_{+,in} f_{in} +
\hat{a}^{\dagger}_{+,in} f^{\ast}_{in} 
=\hat{a}_{+,out} g_{out} +
\hat{a}^{\dagger}_{+,out} g^{\ast}_{out}
\nonumber\\
&&\hat{a}_{-,in} F_{in} +
\hat{a}^{\dagger}_{-,in} F^{\ast}_{in} = \hat{a}_{-,in} G_{in} +
\hat{a}^{\dagger}_{-,in} G^{\ast}_{in}. 
\label{med}
\end{eqnarray}
Using  Eqs. (\ref{transf}) in Eq. (\ref{med}) we obtain that the
creation and the destruction operators after the amplification 
($\hat{a}_{\pm,in}$) are related to the ones
before the amplification ($\hat{a}_{\pm,in}$)
by the following Bogoliubov-Valatin transformations:
\begin{eqnarray}
&&\hat{a}_{+,in} = c_{+} \hat{a}_{+,out} + 
c^{\ast}_{-}\hat{a}^{\dagger}_{+,out},~~~
\label{bog1}\\
&&\hat{a}_{-,in} = \tilde{c}_{+} \hat{a}_{-,out} + 
\tilde{c}^{\ast}_{-}\hat{a}^{\dagger}_{-,out}, ~~~
\label{bog2}
\end{eqnarray}
Once the evolution of the pseudoscalar field $\psi$ is
 completely specified, the
coefficients $c_{\pm}$ and $\tilde{c}_{\pm}$ can be explicitly
computed.
It is useful, at this point to re-write the full expression of the
hypercharge after the amplification took place
\begin{eqnarray}
&&\vec{ Y}_{out}(\vec{x},\tau) = 
\frac{1}{\sqrt{2}}\int\frac{d^3 k} 
{\sqrt{(2\pi)^{3}}}
\Biggl\{\Biggl[\Biggl(\hat{a}_{+,out} g_{out} +
\hat{a}^{\dagger}_{+,out} g^{\ast}_{out}\Biggr) \vec{\epsilon}_{+} 
+ \Biggl( \hat{a}_{-,out} G_{out} +
\hat{a}^{\dagger}_{-,out} G^{\ast}_{out}\Biggr) 
\vec{\epsilon}_{-} \Biggr] e^{i \vec{k}\cdot\vec{x}} 
\nonumber\\
&&+\Biggl[\Biggl(\hat{a}^{\dagger}_{+,out} g^{\ast}_{out} +
\hat{a}_{+,out} g_{out}\Biggr)\vec{\epsilon}_{-} 
+  \Biggl( \hat{a}^{\dagger}_{-,out} G^{\ast}_{out} +
\hat{a}_{-,out} G_{out}\Biggr)\vec{\epsilon}_{+} \Biggr] e^{-i
\vec{k}\cdot\vec{x}}\Biggr\}.
\label{decomposition3}
\end{eqnarray}
We have now all the ingredients in order to estimate the expectation values 
of the pseudoscalar quantities we are interested in. As it was previously 
stressed \cite{m2,m3} a good gauge-invariant measure of global parity 
violation is represented by $\vec{H}_{Y} \cdot \vec{\nabla}\times \vec{H}_{Y}$. 
This operator emerges naturally in the study of the evolution of 
hypermagnetic fields at finite conductivity and finite fermionic density. 
As we will show in the next two sections, the same provides a good
gauge-invariant measure of parity breaking both for $T\geq T_{c}$ and for
 $T\leq T_{c}$.

Then we can compute our expectation value by using Eq. 
(\ref{decomposition3}) and the Bogoliubov transformations 
(in order to relate  incoming and outgoing modes):
\begin{eqnarray}
\langle 0| \vec{H}_{Y} \cdot \vec{\nabla}\times 
\vec{H}_{Y} 
|  0 \rangle = \int \frac{d^3 p}{(2 \pi)^3} p^3 
\biggl[&& (|c_{+}|^2 + |c_{-}|^2) f_{in}(\tau) f_{in}^{\ast}(\tau) - 
(|\tilde{c}_{+}|^2 + |\tilde{c}_{-}|^2) F_{in}(\tau) F_{in}^{\ast}(\tau)
\nonumber\\
&+&
c^{\ast}_{+} c_{-} f^2_{in}(\tau) + c_{+} c^{\ast}_{-} 
{f_{in}^{\ast}}^2(\tau) - 
\tilde{c}^{\ast}_{+} \tilde{c}_{-} F^2_{in}(\tau) - 
\tilde{c}_{+} \tilde{c}^{\ast}_{-} {F_{in}^{\ast}}^2(\tau)\biggr],
\label{expectation}
\end{eqnarray}
where $|0\rangle$ denotes the vacuum state of the hypercharge field.
 Suppose that our initial field configuration is given by a 
stochastic background or by a thermal mixture \cite{mg}. 
Then $f_{in}(\tau)$ and 
$F_{in}(\tau)$  will be 
\begin{equation}
f_{in}(\tau) = \frac{{\cal A}(k)}{\sqrt{k}} e^{- i k \tau}, ~~~F_{in}(\tau) =  
\frac{{\cal A}(k)}{\sqrt{k}} e^{- i (k \tau + \varphi)}. 
\end{equation}
If there is no initial phase shift (as it should be) $\varphi =0$ and then 
the two mode functions are equal. Thus, going to Eq. (\ref{expectation}) 
we have that the whole expression is zero provided $c_{\pm} \neq 
\tilde{c}_{\pm}$. Indeed  we showed that because $Y_{+}$ and $Y_{-}$ 
obey different classical equation the Bogoliubov coefficients with and 
without tilde will be in general different. The specific form of the 
Bogoliubov coefficients will of course depend upon the specific  
cosmological model, upon the specific particle physics candidate for $\psi$.
However, even in the general case we can carry on the calculation one 
step further. The amplification coefficients are 
represented, in our formalism, by $c_{-}$ and $\tilde{c}_{-}$. 
Thus, 
the logarithmic energy 
spectrum of the hypermagnetic helicity can be written as 
\begin{equation}
\frac{d {\cal H}e_{Y}}{d \log{\omega}} = \frac{1}{ \pi^2} \omega^5 
{\cal G}(\omega,\tau) |{\cal A}(\omega)|^2,
\label{hel2}
\end{equation}
with 
\begin{equation}
{\cal G}(\omega,\tau) = \biggl[|c_{-}(\omega)|^2 -  
|\tilde{c}_{-}(\omega)|^2   - 
c^{\ast}_{+} c_{-} f^2_{in}(\tau) + c_{+} c^{\ast}_{-} 
{f_{in}^{\ast}}^2(\tau) - 
\tilde{c}^{\ast}_{+} \tilde{c}_{-} F^2_{in}(\tau) - 
\tilde{c}_{+} \tilde{c}^{\ast}_{-} {F_{in}^{\ast}}^2(\tau)\biggr],
\end{equation}
where we  defined, as usual, the physical frequency $\omega= k/a$. Notice 
that in deriving Eq. (\ref{hel2}) we used the relation $ |c_{+}|^2 - 
|c_{-}|^2=1$ (and analogously for the tilded quantities) which is 
a trivial consequence of the unitarity.

By defining
the logarithmic energy spectrum of the initial distribution of 
hypemagnetic fields  \cite{m3}
\begin{equation}
\rho(\omega) = \frac{d \rho_{Y}}{d \log{\omega}} = \frac{\omega^4}{\pi^2} 
|{\cal A}(\omega)|^2,
\end{equation}
 Eq. (\ref{hel2}) becomes
\begin{equation}
\frac{d {\cal H}e_{Y}}{d \log{\omega}} = \omega 
{\cal G}(\omega,\tau)~\rho(\omega).
\label{hel3}
\end{equation}
Even if the amplification related to the dynamical pseudoscalars 
is of order of $1$, this last expression tells us that the obtained  
helicity can be large {\em provided the initial state is  not 
 the vacuum}. We will see that the result of Eq. (\ref{hel3}) can be 
quite relevant for our applications.

Therefore, in this section we derived a number of results. As previously 
noted \cite{jak}
we obtained that  dynamical pseudoscalars induce an interaction between 
the polarizations of Abelian gauge fields. This coupling has the effect of 
modifying the evolution of the two circular polarizations in such a way that
the magnetic helicity (initially zero) can be dynamically generated.

Up to now our  results  do not take into account the possible 
effect, in the final state, of some effective Ohmic current which can account
(macroscopically) of  the dissipative effects associated with a plasma at 
finite conductivity. 
 This will be an important point in the next two 
sections. On one hand the finite value of the 
conductivity provides a useful (and physicallly motivated) ultraviolet cut-off
for our helicity and energy spectra. On the other hand we will stress that, 
also at finite conductivity, the Chern-Simons density 
can be directly related to 
the helicity.  

\renewcommand{\theequation}{4.\arabic{equation}}
\setcounter{equation}{0}
\section{Hypermagnetic Knots production during an Inflationary Phase}

The ways  our mechanism can be implemented depend upon the 
early history of the Universe. We will firstly consider a simplified model
where  an inflationary phase occurs  $H$.
 The inflationary phase terminates at the time
 $\tau_1$ and the ordinary (radiation-dominated) phase settles in.

Notice that 
 $\psi$ can oscillate both in the inflationary phase and in the
radiation dominated phase.
These two logical possibilities involve two different physical pictures. 
In the case of inflationary production of hypermagnetic helicity 
the oscillations will only be damped by the Universe expansion. If, on the
contrary, $\psi$ oscillates during the radiation dominated phase 
we can expect that the evolution of the hypercharge fields will be damped
by the finite value of the conductivity. In the present and in the 
following Sections we will analyse these two regimes.

Suppose that our (massive) pseudoscalar field evolves during an inflationary 
phase following its equation of motion which we write (in cosmic) time as
\begin{equation}
\ddot{\psi} + 3 H \dot{\psi} + m^2 \psi =0,
\end{equation}
where $H$ is (approximately) constant and it corresponds to the maximal 
(curvature) scale reached during inflation.
In this model we have essentially three parameters. One is $H_{i}/M_{P}$ 
fixing the maximal curvature scale during inflation. From the contribution
of gravitational waves to the Cosmic Microwave Background (CMB) anisotropy 
we know that
\begin{equation} 
\frac{H_{i}}{M_{P}} \laq 10^{-6},
\end{equation}
where $H_{i}$ is the curvature scale at the end of the inflationary phase.
 Moreover, we also know that the EWPT takes place when 
$T>100$ GeV. Since, by that time, the Universe was dominated by 
radiation we also have to assume that 
\begin{equation}
\frac{H_{i}}{M_{P}} > 10^{-33}.
\end{equation}
The second parameter of our model is $m/H_{i}$ which essentially fixes the 
amount of amplification of the hypercharge. 
In Eq. 
(\ref{decoupled}) the instability develops for $k \sim k_{{\rm max}} \sim c 
(\psi_0/M) m a$. The  amplification occurring for $k\simeq
 k_{{\rm max}}$ 
in the mode function will be of the order of 
$Y_{+}(k, \tau) \sim \exp{[c (\psi_0/M) m a \Delta \tau]}$ where 
$ a \Delta \tau \sim \Delta t \sim H_{i}^{-1}$ is the typical amplification 
time. The typical growth of $Y_{+}$ will then be proportional to 
$\exp{[c (\psi_0/M) (m/H_{i})]}$. As we can see the amplification depends 
also upon $(\psi_0/M)$ which is the third parameter of our analysis. 
The coupling constant $c$ appearing in the action (\ref{action}), is in 
principle, determined by the specific particle physics model. We want 
immediately to notice that the amplification occurs for $m \gg H_{i}$. 
At the same time the coherence length of the amplification processes is 
determined by 
$L_{{\rm max}} \sim \omega_{{\rm max}}^{-1} \sim c^{-1} (M/\psi_0) m^{-1}$ 
and, therefore,
the larger the amplification is the shorter is the coherence length of
the amplified field in horizon units.
If $m >H_{i}$ effect of the Universe expansion can be neglected at least in
 the first approximation. 

\begin{figure}
\begin{center}
\begin{tabular}{|c|c|}
      \hline
      \hbox{\epsfxsize = 7 cm  \epsffile{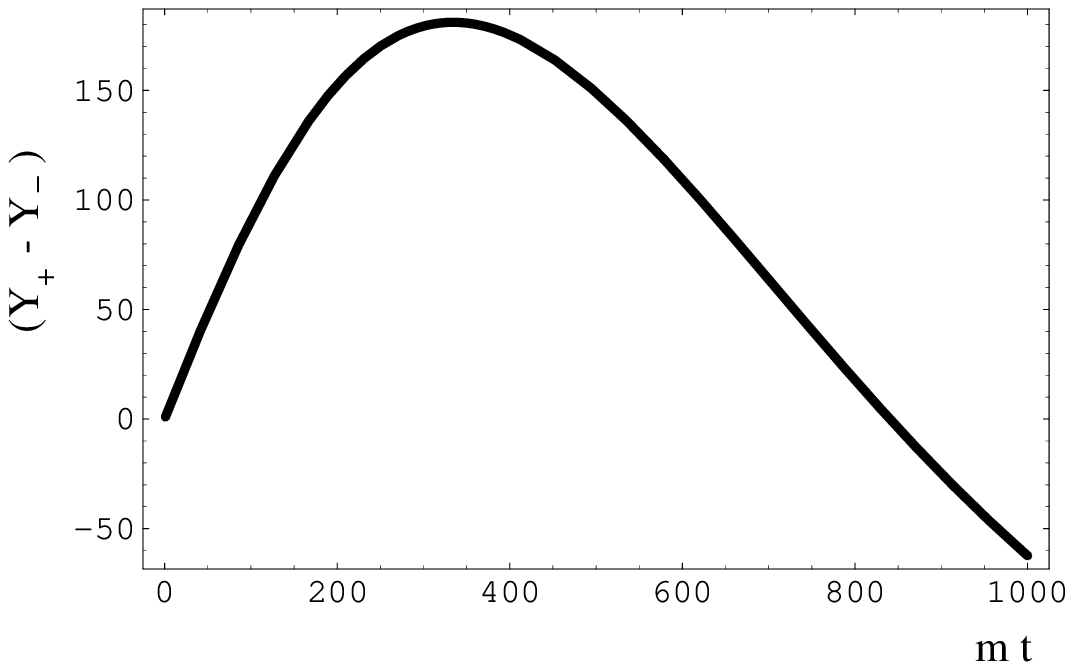}} &
      \hbox{\epsfxsize = 7 cm  \epsffile{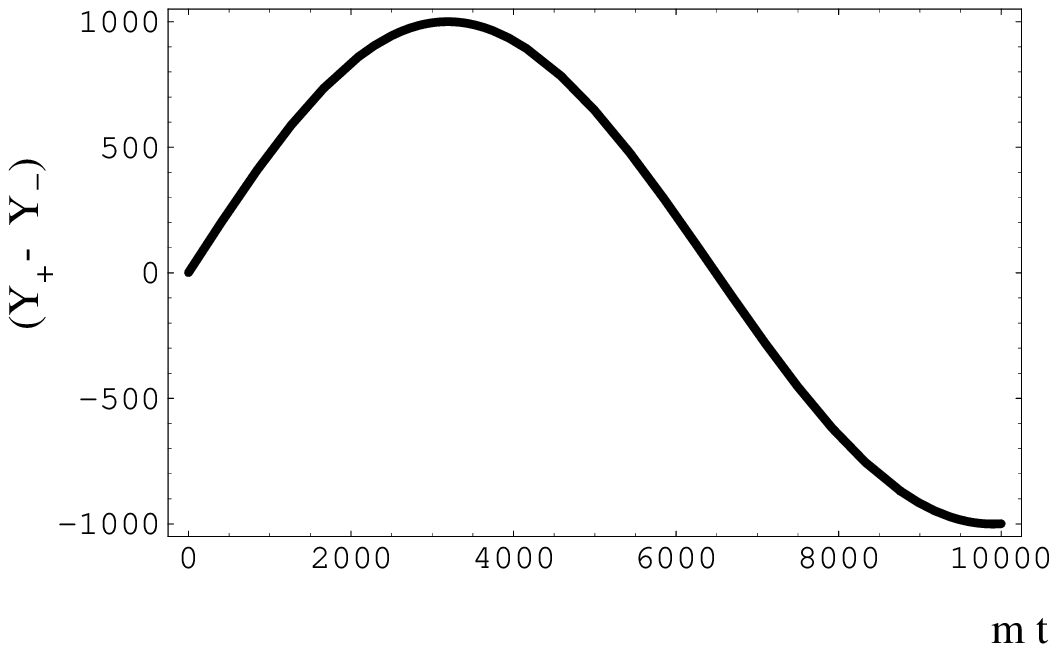}} \\
      \hline
\end{tabular}
\end{center}
\caption[a]{We report two numerical solutions of the system given in 
Eqs. (\ref{cosmic}). We assume $\omega=\omega_{{\rm max}}$. At the
left we report the case of $c= 0.01$ and $H/m = 0.001$. At the right we 
report the case with $c=0.001$ and $H/m \sim 0.00013$. For the two cases we 
would get, from our considerations following Eq. (\ref{flat}), amplifications
 of the order of $e^{5}$ (left) and $e^{8}$ (right). The damping due to 
expansion settles in quickly (roughly after $H^{-1}$ in mass units). The 
agreement our our estimates with the numerical evaluation gets worse for
frequencies far from the maximal amplified frequency.}
\label{num1}
\end{figure}

This qualitative analysis can be refined by a numerical study which includes
the Universe expansion.
For this purpose we can write the hypercharge evolution in cosmic time. The 
system which should be solved numerically is, in general,
\begin{eqnarray}
&&\ddot{y}_{\pm} + \biggl[ \omega^2 \mp \omega \frac{c}{M} \dot{\psi} - 
\frac{H^2}{4} - \frac{\dot{H}}{2} \biggl] y_{\pm} =0,
\nonumber\\
&&\ddot{\Psi} + \biggl[ m^2 - \frac{9}{4} H^2 - \frac{3}{2} \dot{H}\biggl] 
\Psi=0,
\nonumber\\
&&y_{\pm}=\sqrt{a} Y_{\pm},~~~ \Psi = a^{3/2} \psi.
\label{cosmic}
\end{eqnarray}
In our specific case this second order 
differential system can be simplified since $\dot{H}=0$.
We studied numerically this set of equations by reducing it to a first
order linear differential system. Some examples of our
 results are reported in Fig. \ref{num1}. The crucial point in our analysis
is the determination of the maximally ampplified mode 
$\omega_{{\rm max}}$. Once this 
is obtained we can  solve the system given in Eq. (\ref{cosmic}) to include the
Universe expansion. 
In the approximation where $m\gg H$ (i. e. expansion negligible
 compared to the 
oscillations) we have that the Eqs. (\ref{cosmic})
 can be further simplified since $\psi \sim a^{-3/2}
\sin{[m (t - t_0)]}$. We get a Mathieu-type equation 
\cite{abramowitz,diff}
\begin{equation}
\frac{d^2y_{\pm}}{dz^2} + \biggl[ \delta \mp \epsilon \cos{2 z}\biggr]
y_{\pm} =0,
\label{flat}
\end{equation}
where $\delta = (2\omega/m)^2$, $\epsilon = c (\psi_{0}/M) (2\omega/ m)$ 
and $z = (m t)/2$. Note that when $\epsilon=0$ and $\delta=(2 k+ 1)^2$
 (where $k=0,1,2...$) there is 
solution of Eq. (\ref{flat}) with period $2 \pi$ but when $\epsilon=0$ and 
$\delta=(2 k)^2$  there is a solution of Eq. (\ref{flat}) with period $\pi$.
Therefore, in the plane $(\delta,\epsilon)$ the instability boundaries will 
cross the $\epsilon=0$  at the points defined by $\delta= (2 k)^2$ or 
$\delta= (2 k+1)^2$. By studying the instability of Eq. (\ref{flat}) in the 
case of $\epsilon\neq 0$ (but still $\epsilon <1$ ) we get that the first
instability region (corresponding to $\delta\sim 1$) occurs for a region in the 
plane $(\delta,\epsilon)$ bounded by $\delta- 1 \sim \pm \epsilon/2$. 
Inserting the 
values of $\delta$ and $\epsilon$ we get that, 
for small $\epsilon$ this region 
is centered around $\omega_{{\rm max}} \sim (c/2) (\psi_{0}/M) m$. Of course 
there will be also  other bands defined by their intersections on the 
$\epsilon=0$ axis ($\delta=4,~9,~16,~20...$). Those bands correspond to higher
frequencies but they are narrower than the first one defined by the 
$\delta=1$ intersection \cite{abramowitz,diff}. 

According to this analysis we can say that $\omega_{{\rm max}}$ leads 
to an amplification of the order of $\exp{[ (c/2) (\psi_0/M) m \Delta t]}$, 
where
$\Delta t\sim H^{-1}$. As noted in previous studies of a 
similar equation,  the Universe expansion will
soon shut-off the instability occurring in the $\epsilon<1$ case \cite{reh}. 

In Fig. \ref{num1} we report 
the solution of the system given by Eq. (\ref{cosmic}) where we took  
$\omega=\omega_{{\rm max}}$. The Universe 
expansion has been included. If we take the case with  $c= 0.01$ and 
$m/H=1000$ we  get that the amplification should be $e^{5}= 148$. 
Numerically we get that during one Hubble time the maximal amplification
is roughly $150$ (left picture in Fig. \ref{num1}). However the 
amplification is quickly shadowed by the Universe expansion which becomes
effective roughly after a time of the order of $H^{-1}$ in mass units.

Thus  pseudoscalar oscillations are quite effective
in changing the topology of a stochastic background by inducing 
correlations between the two independent polarizations of the gauge field. 
However, they are not effective in {\em inflating} the amplitude of
 the background.
 Therefore, as we will see even better in the case of oscillations 
occurring in a relativistic plasma, 
one of the crucial assumption of our considerations will be the existence 
of a primordial stochastic background of hypermagnetic fields whose topology
can be eventually twisted by the pseudoscalar oscillations.

In Fig. \ref{num1} we 
completely ignored the role played by the finite value of the conductivity.
During the inflationary dominated regime the effective 
conductivity is equal to zero since no charged particles are 
present in the Universe. When the Universe reheats charged particles 
are generated and, therefore, an Ohmic current is developed. This 
effect can be modeled by thinking about the situation where the 
conductivity is zero prior to the onset of the radiation dominated phase
but it gets larger and larger as soon as charged particles are created.

Needless to say that the the smooth transition of the conductivity from 
zero to a finite (large) value is quite important for our considerations.
Suppose, as a warm-up, that we are in flat space and suppose that 
the conductivity jumps  from zero to a finite value at some 
time $t_{\star}$. Let us 
also assume, for simplicity, that there are no pseudoscalar interactions 
in the game.  Then the 
evolution equation of the hypercharge fields will be the usual D'Alembert 
equation for $t<~t_{\star}$ but it will be modified as: 
\begin{equation}
\vec{{\cal Y}}'' + \sigma_c \vec{{\cal Y}}' - \nabla^2 \vec{{\cal Y}}=0,
\end{equation}
for $t>~t_{\star}$. If we match the two solution at $t=t_{\star}$ we 
simply get that the hypermagnetic fields are left unchanged by the transition
whereas the hyperelectric fields are damped  \cite{mof}:
\begin{equation}
{\vec {\cal H}}_{>} (t_{\star}) ={\vec {\cal H}}_{<} (t_{\star}),~~~
{\vec {\cal E}}_{>} (t_{\star}) =e^{- \sigma_c t_{\star}}
{\vec {\cal E}}_{<} (t_{\star}),
\end{equation}
where the subscript $>$ ($<$) denotes the value of the fields after (before)
the transition occurring at $t_{\star}$. Since pseudoscalar  interactions 
are absent in this example we also have that $({\cal Y}_{+} -{\cal Y}_{-})$
will always be zero.
 
\begin{figure}
      \centerline{\epsfxsize = 8 cm  \epsffile{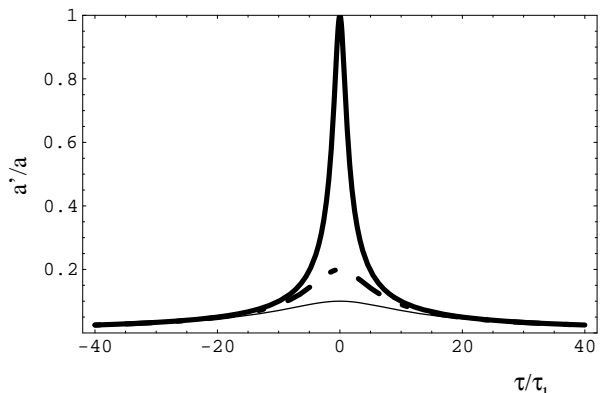}} 
\caption[a]{We illustrate the Hubble parameter ${\cal H} = a'/a$ computed
from Eq. (\ref{scalef}). By increasing $\tau_1$ the transition regime 
gets longer. For $\tau_1= 1$ the Hubble parameter is reported 
with the full (black) line. For larger values of $\tau_1$ 
(i.e. $\tau_1 =5$ [ thick dashed line] and $\tau_1= 10$ [full thin line]) the
height of the Hubble parameter decreases and the transition regime gets 
broader.}
\label{reh1}
\end{figure}

From a physical point of view we can expect that the effect of the 
finite value of the conductivity will be to fix a definite sign 
for $(Y_{+} -Y_{-})$. By ignoring the effect of the conductivity we can argue
(see Fig. \ref{num1}) that $(Y_{+} - Y_{-})$ is certainly amplified but, 
at the same time, the  {\em sign} of this quantity is not well defined.
In order to clarify this point let us study the full system of equations 
at finite conductivity
\begin{eqnarray} 
&&Y_{\pm}'' + \sigma Y_{\pm}'+ \biggl[ k^2 \mp k \frac{c}{M} \psi'\biggr]
 Y_{\pm}=0,
\nonumber\\
&&\psi'' + 2 {\cal  H} \psi' + m^2 a^2 \psi=0,
\label{conduc}
\end{eqnarray}
where we remind that the conductivity, in curved space, evolves as 
$\sigma(\tau) = \sigma_c a(\tau)$.
In order to  model a smooth transition from the inflationary epoch to the 
radiation dominated epoch let us assume that the geometry evolves
continuously from a de Sitter (or quasi de Sitter) stage to a radiation 
dominated stage of expansion.
This can be achieved with a specific choice of the scale 
factor, namely
\begin{equation}
a(\tau) = \biggl(\frac{\tau}{\tau_{1}}\biggr) + 
\sqrt{\biggl(\frac{\tau}{\tau_1}\biggr)^2 + 1}.
\label{scalef}
\end{equation}
Notice that for $\tau \rightarrow +\infty$ we have that $ a(\tau) \sim \tau$. 
For $\tau\rightarrow -\infty$ we have that $a(\tau) \sim \tau^{-1} $. 
For $\tau\rightarrow 0$ we have that $a(\tau) \sim {\rm constant}$. So
we can say that this geometry evolves smoothly from a de Sitter phase 
to a radiation dominated phase. The Hubble parameter in conformal 
time can be easily computed and it is also a smooth function of $\tau$
\begin{equation}
{\cal H}(\tau) \equiv \frac{ a'}{a} = \frac{1}{\sqrt{\tau^2 + \tau_{1}^2}}.
\end{equation}
The duration of the transition period between the de Sitter phase and the 
radiation dominated phase is controled by $\tau_1$. By increasing $\tau_1$ the 
transition period gets longer. By decreasing $\tau_1$ the transition period
 gets shorter. This aspect can be appreciated by
 looking at the time evolution of ${\cal H}(\tau)$ (see Fig. \ref{reh1}). 
The behavior of the conductivity follows from the evolution of the geometry. 
Therefore, in this model, we will have that the conductivity is vanishingly 
small during the de Sitter phase and it grows linearly in conformal 
time during the radiation dominated phase. By rescaling 
with respect to the mass in order to have fully dimensionless quantities 
we can easily re-write Eqs. (\ref{conduc}) in the form of  a first order differential system
\begin{eqnarray}
&& \frac{d q_{+}}{d x}= p_{+}(x),~~~\frac{d q_{-}}{d x}= p_{-}(x),~~~
\frac{d u}{d x} = v(x),
\nonumber\\
&&\frac{d p_{+}}{d x} = - \biggl(\frac{\sigma_{c}}{m}\biggr) a(x) p_{+} -
\biggl[ \varepsilon^2 - \varepsilon \frac{c}{M} u(x)\biggr] q_{+}(x),
\nonumber\\
&&\frac{d p_{-}}{d x} = - \biggl(\frac{\sigma_{c}}{m}\biggr) a(x) p_{-} -
\biggl[ \varepsilon^2 + \varepsilon \frac{c}{M} u(x)\biggr] q_{-}(x),
\nonumber\\
&& \frac{d v}{d x} = -  \frac{2~v(x)}{\sqrt{ 
x^2 + x_1^2}} - a^2(x) u(x),
\label{conduc2}
\end{eqnarray}
where we denoted  $q_{\pm} \equiv Y_{\pm} $, $ p_{\pm} = Y_{\pm}'$ and 
$u = \psi$ and $v= \psi'$. Notice, moreover that $ x = m \tau$,
$x_1 = m \tau_{1} $ and $\varepsilon = k/m$.  
We can  observe that there are three
 important parameters
for our discussion. The first one is, 
${\cal H}_{1} = {\cal H}(\tau_1)\sim \tau_1^{-1}$
  the Hubble parameter at the end of the inflationary phase. The Hubble 
parameter in cosmic time is  simply related to ${\cal H}_1$ since 
${\cal H}_1 = a_1 H_1\sim (1 + \sqrt{2}) H_{1}$. Moreover, as we noticed
 before, in order to be compatible with the large scale observations 
in the microwave sky we have to require $ H_{1}/M_{P} < 10^{-6}$.
 Notice that  $H_1$ always 
appears in the combination $x_1 = m \tau_1 \sim m/H_1$. 
Consequently,  if we want large  pseudoscalar oscillations prior to 
$\tau_1$ we have to require $x_1 \ll 1$. The last parameter appearing 
in Eqs. (\ref{conduc2}) is $ \sigma_c/m$. Numerically
$\sigma_c= \sigma_0 T_0$ where $\sigma_0 = 70$--$100$ and $T_0$ is the 
temperature of the Universe at the beginning of the 
radiation dominated epoch. Notice that $\sigma_c/m \equiv (\sigma_{c}/H_1)
(H_1/m)$. Therefore, taking into account that $H_1 < 10^{-6} M_{P}$ and 
that we want $T_{0}$ as high as $10^{10}$ GeV we have to require that 
$\sigma_{c}/m<1$. Suppose, for instance, that $H_1\sim 10^{-9} M_{P}$,
$T_0\sim 10^{10}$ GeV, $m/H_1 =100$. Then, we get that 
$\sigma_c/m= \sigma_0/100 \sim 0.7$. We will keep $x_1$ and $ \sigma_c/m$ as
 free parameters in the integration. Notice finally that we will 
focus our attention on the maximally amplified modes and, therefore, we will
fix $\varepsilon\sim c/2$. 
\begin{figure}
\begin{center}
\begin{tabular}{|c|c|}
      \hline
      \hbox{\epsfxsize = 7 cm  \epsffile{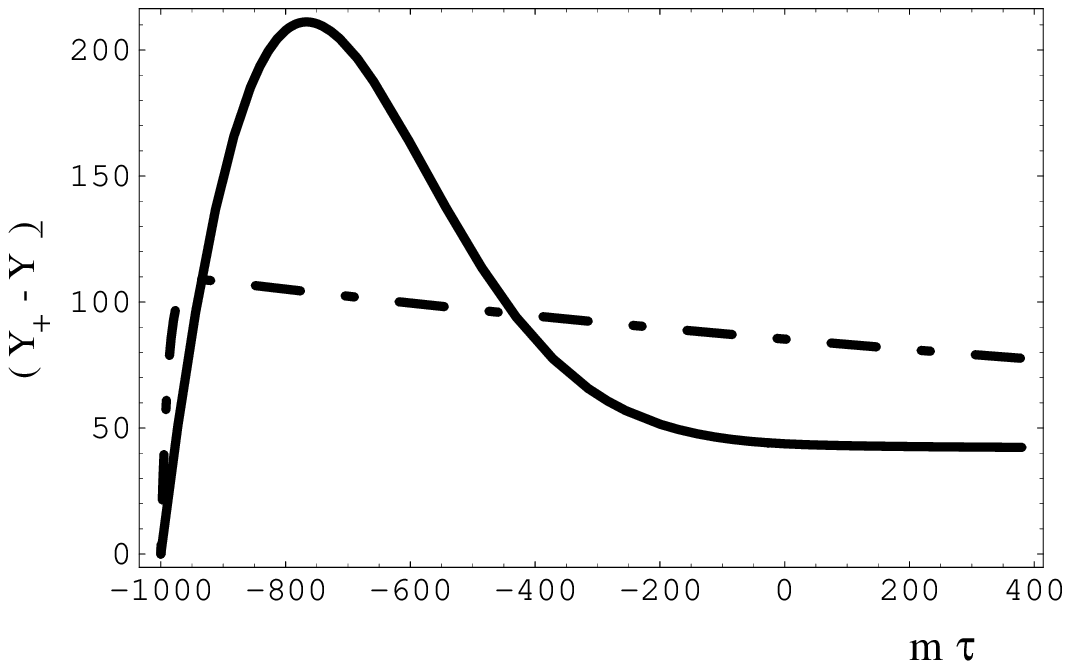}} &
      \hbox{\epsfxsize = 7 cm  \epsffile{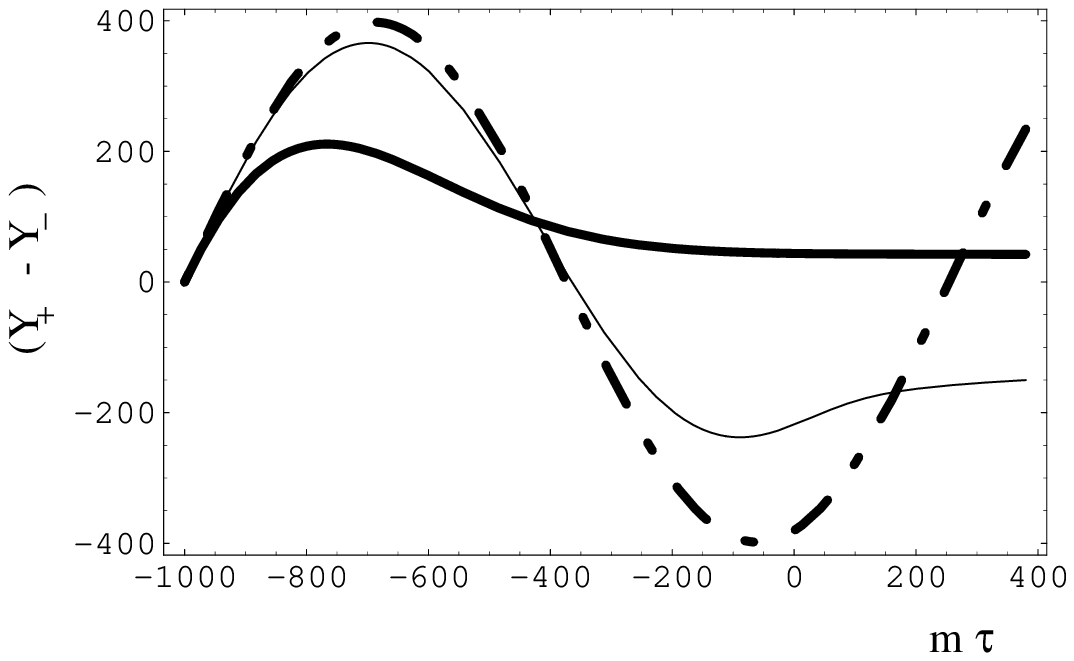}} \\
      \hline
\end{tabular}
\end{center}
\caption[a]{We illustrate the numerical integration of the system given in
Eqs. (\ref{conduc2}). At the left we report the cases where 
$\sigma_c/m=0.1$ and $x_1=100$ (full thick line) and $x_1= 10^{4}$ 
(dot-dashed line). At the right we fix $x_1=100$ and we choose, respectively,
$\sigma_c/m =0.1$ (full thick line), $\sigma_c/m=0.01$ (full thin line) and 
$\sigma_c/m = 10^{-4}$ (dot-dashed line).}
\label{reh2}
\end{figure}
In Fig. \ref{reh2} and \ref{reh3} we report the results of the numerical 
integration for different sets of parameters.
In Fig. \ref{reh2} (left plot) we fix $\sigma_c/m\sim 0.1$ and we 
choose $x_1=100$ (full line), $x_1=10^{4}$. We see that $(Y_{+} -Y_{-})$ 
acquires a definite sign as we expected. Moreover the transition regime 
gets broader as soon as we increase $x_1$. This can be understood 
by recalling that by increasing $\tau_1$ the transition regime gets more 
significant.
What happens if we change $\sigma_c/m$? In Fig. \ref{reh2} (left plot)
we fix $x_1=100$ and we decrease the value of $\sigma_c/m$. The the 
full (thick) line corresponds to the case $\sigma_c/m=0.1$ , whereas the 
dot-dashed line corresponds to the case $\sigma_{c}/m = 10^{-4}$. 
The full (thin) line corresponds to the case $\sigma_{c}/m=0.01$.
We clearly
see that in the limit $\sigma_c/m\rightarrow 0$ we recover 
partially the case of Fig. \ref{num1} since $(Y_{+} - Y_{-})$ does not 
have definite sign when the radiation dominated phase settles in.
The fact that $\sigma_{c}/m$ cannot be too small can be understood physically 
by recalling that the small $\sigma_{c}/m$ limit is equivalent to a small
 reheating temperature. What happens when $\sigma_{c}/m >1$.  
We can expect that if the conductivity is too large at the beginning of the 
radiation dominated epoch the damping term will be dominant and the 
amplification will be reduced. This aspect is illustrated in Fig. \ref{reh3}.
\begin{figure}
      \centerline{\epsfxsize = 7 cm  \epsffile{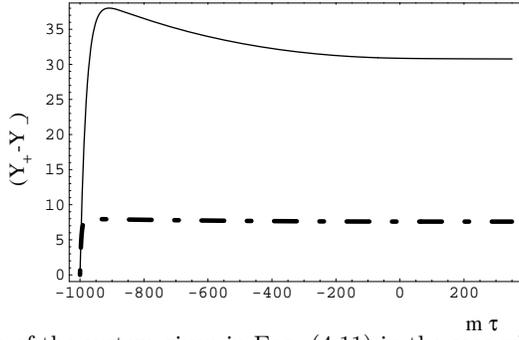}} 
\caption[a]{We illustrate the integration of the system given in Eqs. 
(\ref{conduc2}) in the case where $x_1= 100$, $\sigma_c/m = 1$ 
(full thin line) and in the case where $x_1= 100$, 
$\sigma_c/m = 5$ (dot-dashed line).}
\label{reh3}
\end{figure}
The limit $\sigma_{c}/m\gg 1$ simply means that we are in the damped 
regime which will be specifically addressed in Sections V and VII.

In the sudden approximation, the Bogoliubov coefficients describing the 
amplification of the mode function when the background changes
from the inflationary phase to the radiation dominate phase can be 
analytically computed in the framework of our model.
When the modes of the fields evolve in the radiation dominated 
phase we will also have to take into account the effect of the 
conductivity and this  will be the second step of our 
calculation. 
We will solve the evolution of the mode function for modes
close to $\omega_{{\rm max}}$. 
For $\tau <\tau_0$ the pseudoscalar does not oscillate but when its mass
is of the order of $H_{i}$ the inflationary oscillations will begin until
the radiation dominated phase starts at $\tau\sim \tau_{1}$. 
The mode function corresponding to $Y_{+}$ for
 $k \geq k_{{\rm max}}$ evolves in the three regions as
\begin{eqnarray}
g_{I}(\tau) &=& \frac{1}{\sqrt{k}}e^{-i k\tau},~~~~~~~~~~~~~~\tau<\tau_{0},
\nonumber\\
g_{II}(\tau) &=&\frac{1}{\sqrt{k}}\biggl[ B_{+} e^{\beta\tau}
+ B_{-} e^{-\beta\tau}\biggr], ~~~~\tau_0 < \tau<\tau_{1},
\nonumber\\
g_{III}(\tau) &=& \frac{1}{\sqrt{k}}\biggl[ c_{+} e^{-i k\tau} + c_{-},
e^{i k \tau}\biggr], ~~~~\tau >\tau_1,
\label{sol1}
\end{eqnarray}
where $\beta = c/2 (\psi_0/M) m a$. We want to stress that the solution
for $\tau_{0} <\tau< \tau_1$ holds for modes $k\sim k_{{\rm max}}$ and for
times $a\Delta\tau \laq H^{-1}$ (where $\Delta\tau \sim \tau_{1} -\tau_0$)
but it should not be trusted outside of this range as discussed
 in Fig. \ref{num1}. 
Notice that $c_{\pm}$ are exactly the Bogoliubov coefficients we are
 looking for since they relate the incoming mode function with the
outgoing mode function. They can be determined by matching the 
three solutions (and their first derivatives) in
the three regions defined by $\tau_0$ and $\tau_1$. This can be easily done
and the result is
\begin{eqnarray}
&&c_{+}= \frac{1}{2} e^{-i k\Delta\tau } \biggl[ 2
\cosh{[\beta\Delta\tau]} - i \biggl( \frac{k}{\beta} -
\frac{\beta}{k}\biggr) \sinh{[\beta\Delta\tau]} \biggr],
\nonumber\\
&&c_{-}=- \frac{i}{2} e^{-i k\Delta\tau}\biggr[
\biggl(\frac{k}{\beta} + \frac{\beta}{k} \biggr)
\sinh{[\beta \Delta\tau]}\biggr],
\label{c1}
\end{eqnarray}
where $\Delta\tau= \tau_{i} - \tau_0$ defines the duration 
of the oscillatory phase. During the time the $Y_{+}$ is 
amplified we also have that the $Y_{-}$ will oscillate 
and the corresponding Bogoliubov coefficients will
not describe amplification but mixing between the 
positive and negative frequencies. They are 
\begin{eqnarray}
&&\tilde{c}_{+}= \frac{1}{2} e^{i k\Delta\tau } \biggl[ 2
\cos{[\beta\Delta\tau]} + i \biggl( \frac{k}{\beta} +
\frac{\beta}{k}\biggr)\sin{[\beta\Delta\tau]} \biggr],
\nonumber\\
&&\tilde{c}_{-}= -\frac{i}{2} e^{-i k\Delta\tau}\biggr[
\biggl(\frac{k}{\beta} - \frac{\beta}{k} \biggr)
\sin{[\beta\Delta\tau]}\biggr].
\label{c2}
\end{eqnarray}
Our expressions of the Bogoliubov coefficients
satisfy $|c_{+}|^2 - |c_{-}|^2 =1$, and 
$|\tilde{c}_{+}|^2 - |\tilde{c}_{-}|^2 =1$ as required by the
unitarity of the Bogoliubov transformation which preserves 
the commutation relations between the creation and annihilation
operators in the asymptotic regions. 

According to Eq. (\ref{hel2}),
 the mean helicity spectrum in the case ${\cal A}(k)=1$ will be:
\begin{equation}
\frac{d {\cal H}e_{Y}}{d \log{\omega}} \simeq \frac{\omega_{{\rm max}}^5}{\pi^2} 
e^{ c (m/H_{i}) (\psi_{0}/M)}.
\label{helinfl}
\end{equation}
Thus according to Eq. (\ref{hel3}) we will have:
\begin{equation}
\frac{d {\cal H}e_{Y}}{d \log{\omega}} \simeq
\frac{c ~m}{2}\biggl(\frac{\psi_0}{M}\biggr) e^{ c (m/H_{i})
 (\psi_{0}/M)}\rho(\omega_{{\rm max}}),
~~~\rho(\omega_{{\rm max}}) =\frac{1}{\pi^2}(\omega_{{\rm max}})^4
 |{\cal A}(\omega_{{\rm max}})|^2 .
\end{equation}
In Eq. (\ref{helinfl}) we assumed that the hypermagnetic field 
modes are re-entering in the radiation dominated epoch but 
we did not take into account properly the damping 
effect of the conductivity. If the field modes are amplified 
(as in our case) during a phase where the conductivity is strictly zero
(since no charged particles are present) the only effect of the 
conductivity (arising when the field modes enter in the radiation dominated 
phase) is to cut-off (exponentially) all the momenta larger than the typical 
momentum associated with the conductivity. However, if the 
pseudoscalar oscillations occur in the radiation dominated phase 
the effect of the conductivity can be even more important, as we will
show in the next section. 

\renewcommand{\theequation}{5.\arabic{equation}}
\setcounter{equation}{0}
\section{Hypermagnetic Knots production during a Radiation Phase}

If pseudoscalar particles are present prior to  the onset of the electroweak
 epoch they can offer a mechanism for the production of Chern-Simons 
condensates whose decay can seed the BAU. An example of such a particle 
is the axial Higgs whose mass enters, after supersymmetry breaking, in the 
expression of the neutral Higgs boson mass \cite{hab,axh}.
Inspired by the  axial Higgs  we can require the 
pseudoscalar mass to be larger than $300$ GeV 
\footnote{Assuming no invisible decays, the direct 
limit is $m > 65$ GeV (at $95 \%$ confidence level and in the case 
$\tan\beta>1$)\cite{PDG}.}. If $ m > 300$ GeV  the 
MSSM Higgs sector is not too different from the one of the MSM.

We want to extend the discussion of hypermagnetic helicity generation to 
the case of a radiation dominated epoch.
Suppose now that at some higher curvature scale a stochastic distribution of 
hypermagnetic field has been created. Then, suppose that at some
lower energy scale the pseudoscalar oscillations will start  developing. 
In this case the evolution of the hypercharge fields is 
\begin{equation}
Y''_{\pm}+ \sigma Y_{\pm}' 
+ \biggl[  k^2 \mp k c \frac{\psi'}{M}\biggr]Y_{\pm} =0.
\label{con1}
\end{equation}
By eliminating 
the first derivative through a field rescaling we get 
\begin{equation}
y_{\pm}'' + \biggl[ k^2 \mp k \frac{c}{M}\psi' - \frac{\sigma^2}{4} -
 \frac{\sigma'}{2}\biggr] y_{\pm} =0,~~~
Y_{\pm}= y_{\pm} \exp{[ - \frac{1}{2} \int^{\tau} \sigma(\tau') d\tau']}.
\label{con2}
\end{equation}
If an initial (stochastic) 
hypermagnetic field is present, then the pseudoscalar interactions will
be able to twist the topologically trivial distribution. If 
no initial stochastic 
distribution is present, the obtained hypermagnetic helicity will be quite 
small. In other words we find that at finite conductivity the pseudoscalar
interactions are only  efficient in twisting a typologically 
trivial hypermagnetic distribution. Indeed,
as we previously discussed, we can always define a conductivity scale at any 
time in the radiation dominated epoch. All the modes of the fields with
typical frequency larger than the typical conductivity frequency
 are essentially
washed away whereas the other have some chance of being amplified.
Eq. (\ref{con2}) becomes, in the large conductivity limit,
 \begin{equation}
\sigma Y_{\pm}' + \biggl[  k^2 \mp k c \frac{\psi'}{M}\biggr]Y_{\pm} =0,
\label{coneq1}
\end{equation}
and the corresponding solution can  be written as
\begin{equation}
Y_{\pm}\sim e^{- \int \frac{k^2}{\sigma(\tau')} d \tau'}\biggl[ e^{\mp \int 
\frac{k}{\sigma(\tau)} \frac{c}{M} \psi'(\tau') d\tau'}\biggr] {\cal A}(\omega)
\label{exp}
\end{equation}
where the initial amplitude ${\cal A}(\omega)$ coincides with the typical 
amplitude of the stochastic hypermagnetic distribution.
The maximal value of the hypermagnetic helicity in critical units can be 
obtained from Eq. (\ref{exp}) with the result that, 
\begin{equation}
\frac{\langle\vec{H}_{Y} \cdot \vec{\nabla}
\times \vec{H}_{Y}\rangle}{\sigma\rho_c}\sim \frac{c }{2 \sigma_0} 
\frac{\Delta\psi}{M} \frac{T_c}{M_0} r(\omega_{{\rm max}}),
\label{con3}
\end{equation}
where $\sigma_0 = \sigma_{c}/T_c$ and $\sigma = \sigma_{c} a$ and $\Delta\psi$ 
is the increment of $\psi$ from its initial value. Typically (without 
fine-tuning the initial amplitude of $\psi$ at the beginning of the 
oscillations) we will have that $\Delta\psi\sim M$. Notice that 
in Eq. (\ref{con3}) we also defined 
\begin{equation}
r(\omega_{{\rm max}}) = \frac{\langle H_{Y}^2 \rangle}{\rho_c} = 
\frac{\omega_{{\rm max}}^4}{\pi^2 \rho_{c}} |{\cal A}(\omega_{{\rm max}})|^2
,~~~~\omega_{{\rm max}} = \frac{c}{2 a} \frac{\Delta\psi}{M} 
\frac{T_{c}^2}{M_0},
\end{equation}
as the critical fraction of energy density present in the form of
a stochastic background of hypermagnetic  fields for 
$\omega\sim \omega_{{\rm max}}$. 
Eq. (\ref{con3}) holds for frequencies smaller than 
 typical conductivity frequency
\begin{equation}
\omega_{\sigma}(\tau)\sim  \sqrt{\sigma_0\sqrt{N_{eff}}}
\sqrt{\frac{T}{M_{P}}} T
\end{equation} 
It is quite relevant to notice that in Eq. (\ref{con3}) 
the exponential damping holds for all the modes $\omega>\omega_{\sigma}$. 
The only way of getting 
a huge amplification (at finite conductivity) would be to require 
$\omega_{\max}\gg \omega_{\sigma}$. However, in this limit, the conductivity 
erases all the field modes. Therefore we are forced to discuss  the case 
$\omega_{{\rm max}}<\omega_{\sigma}$. This limitation on the
amplified physical scales applies for {\em all the 
temperatures (in the symmetric phase of the electroweak theory)
where helicity generation takes place}.

If the conductivity is finite it does not make sense to consider
the case of amplification of the hypermagnetic helicity from 
tha vacuum fluctuations. The reason is that, in this case, 
$r(\omega_{{\rm max}})$ 
is extremely minute. This can be easily seen in general. 
Suppose, in fact that $\omega_{{\rm max}}$ can be made as large as 
$0.1\omega_{\sigma}$. Then we get that the vacuum fluctuations will be
 in this case
\begin{equation}
r(\omega_{{\rm max}}) \sim \frac{10^{-2} \sigma_0}{\pi^2}
 \biggl(\frac{T}{M_{P}}
\biggr)^2
\end{equation}
which is very small for in the symmetric phase of the 
electroweak theory and in the TeV range.

\renewcommand{\theequation}{6.\arabic{equation}}
\setcounter{equation}{0}
\section{BAU from Hypermagnetic Helicity}

We are interested in the possible effects of 
the hypermagnetic fields on the electroweak scale and then we will
mainly consider the plasma effects associated with the dynamics
of Abelian gauge fields in the symmetric phase of the electroweak 
theory. This is consistent with the assumptions made in the 
previous section where we focused our attention on hypermagnetic fields 
amplified and/or re-entered during a radiation dominated phase.

The description of the plasma effects arising in the symmetric 
phase of the electroweak theory has to rely on some generalization
of the MHD equations. In \cite{m3} it was argued that such a 
generalization can be provided by the equations of anomalous 
magnetohydrodynamics (AMHD) where the effect of finite conductivity
and finite chemical potential can be simultaneously described. In the 
present analysis we have to generalize AMHD to the case of pseudoscalar
interactions and non vanishing bulk velocity of the plasma.
Our variables are the  hypermagnetic and
 hyperelectric fields, the right electron chemical potential and the 
pseudoscalar field  $\psi$. Of course these equations have to be supplemented
by the evolution equation for the bulk velocity of the plasma. 
The complete set of equations reads 
\begin{eqnarray}
& &[(p+ \rho) \vec{v}]' + \vec{v}\cdot\vec{\nabla}[(p+\rho)\vec{v}] + 
\vec{v} \vec{\nabla}\cdot[(p+ \rho)\vec{v}] = - \vec{\nabla}p 
+ \vec{J}_{Y} \times \vec{H}_{Y} + \eta [  \nabla^2 \vec{v} + 
\frac{1}{3} \vec{\nabla}(\vec{\nabla}\cdot\vec{v})],~~\eta=(p+\rho)\nu
\nonumber\\
& &{{\vec{H}}_{Y}}' = -\vec{\nabla}
\times {\vec{E}}_{Y},~~~{\vec{\nabla}}\cdot{\vec{H}}_{Y}=0,
\nonumber\\
& &{{\vec{E}}_{Y}}'+ {\vec{J}}_{Y} =  \frac{g'^2}{\pi^2}\mu_{R} a
{\vec{H}}_{Y}+\frac{c}{M} [ \psi' {\vec{H}}_{Y} + \vec{\nabla}\psi\times 
\vec{E}_{Y}] + {\vec{\nabla}}\times{ \vec{H}}_{Y},
\label{mx1}\\
& &{\vec{\nabla}}\cdot {\vec{E}}_{Y}=\frac{c}{M} \vec{\nabla}\psi 
\cdot {\vec{H}}_{Y},~~~
{\vec{J}}_{Y}=\sigma ({\vec{E}}_{Y} + \vec{v} \times \vec{H}_{Y}),
\label{ohm}\\
& & \psi'' + 3 {\cal H}\psi' + m^2 a^2 \psi - \nabla^2 \psi=
-\frac{c}{M} \frac{\vec{E}_{Y} \cdot \vec{H}_{Y}}{a^2},~~~\sigma= a \sigma_{c},
\nonumber\\
& &\frac{1}{a} \frac{\partial (\mu_{R} a)}{\partial \tau} = - \frac{g'^2
}{4\pi^2} \frac{783 }{88} \frac{ 1}{\sigma a^3 T^3}
{\vec{H}}_{Y}\cdot{\nabla\times\vec{H}}_{Y} -
(\Gamma +\Gamma_{H}) (\mu_{R} a) + D_{R} \nabla^2 \mu_{R}, ~~~~\Gamma_{H} = 
\frac{783}{22} \frac{{\alpha'}^2}{\sigma\pi^2} 
\frac{|\vec{H}_{Y}|}{a^3 T^3},
\label{mu}\\
& & \frac{\partial \rho}{\partial \tau} + \vec{\nabla}\cdot[(p+\rho)\vec{v}]
=\vec{E}_{Y}\cdot\vec{J}_{Y},
\label{general} 
\end{eqnarray}
where $\vec{v}$ is the bulk velocity of the plasma, 
$\rho=a^4 \rho_{r}$ and $p=a^4 p_{r}$ are the (rescaled)
energy and pressure densities of the radiation dominated fluid. Notice that 
$\nu$ and  $1/\sigma$ are the thermal and magnetic  diffusivity coefficients
 whereas $D_{R}$ accounts for the dilution of right electrons in the case 
where the chemical potential is not completely homogeneous. 

Using the fact that $\vec{J}_{Y} \sim \vec{\nabla}\times \vec{H}_{Y}$ the 
Navier-Stokes equation, for scales larger than the thermal 
diffusivity scale, becomes 
\begin{equation}
[(p+ \rho) \vec{v}]' + \vec{v}\cdot\vec{\nabla}[(p+\rho)\vec{v}] + 
\vec{v} \vec{\nabla}\cdot[(p+ \rho)\vec{v}] = - \vec{\nabla}[p + 
\frac{1}{2} |\vec{H}_{Y}|^2] 
+ [\vec{H}_{Y}\cdot\vec{\nabla}]\vec{H}_{Y},
\label{ns}
\end{equation}
where we used the fact that $\vec{\nabla}\times\vec{H}_{Y}\times \vec{H}_{Y} = 
- \frac{1}{2}\vec{\nabla} |\vec{H}_{Y}|^2
 + [\vec{H}_{Y}\cdot\vec{\nabla}] \vec{H}_{Y}$.
Notice that we will be always concerned with the case where
$p=\rho/3\gg |\vec{H}_{Y}|^2$ (i.e. vanishing hypermagnetic pressure).
Similarly, the continuity equation will be modified as 
\begin{equation}
\frac{\partial \rho}{\partial \tau} + \vec{\nabla}\cdot[(p+\rho)\vec{v}]
=\frac{1}{\sigma} [\vec{\nabla}\times \vec{H}_{Y}]^2.
\end{equation}
Since we are dealing with modes of the hypercharge field with momentum $k$ 
smaller than
\begin{equation}
k_{\sigma} \sim \sqrt{\frac{\sigma_{c}}{M_{0}}} T, ~~~~M_{0} 
= \frac{M_{Pl}}{1.66 \sqrt{N_{eff}}}\sim 7.1~\times 10^{17} {\rm GeV},
\end{equation}
the source term appearing in the continuity equation can be ignored and
we can indeed 
assume that our plasma is incompressible (i.e.  $\vec{\nabla}\cdot\vec{v}=0$).
In this case, consistency with the continuity equation requires that 
$\rho$ and $p$ depend only upon $\tau$. Then, Eq. (\ref{ns}) can be further 
simplified
\begin{equation}
\frac{\partial\vec{v}}{\partial\tau} + [\vec{v}\cdot\vec{\nabla}]\vec{v} =
\frac{[\vec{H}_{Y}\cdot\vec{\nabla}]\vec{H}_{Y}}{[\rho + p]}.
\end{equation}
With the same set of assumptions we can also obtain a generalized version of
 the hypermagnetic diffusivity equation, namely
\begin{equation}
\frac{\partial{{\vec{H}}_{Y}}}{\partial \tau} =- \frac{4 a
\alpha'}{\pi\sigma}  
\vec{\nabla}\times\left({\mu_{R}\vec{H}}_{Y}\right) -
\frac{c}{M} \vec{\nabla}\times [ \psi' {\vec{H}}_{Y} ]
+{\vec{\nabla}}\times(\vec{v}\times{\vec{H}})
+ \frac{1}{\sigma}
 \nabla^2 {\vec{H}}_{Y}.
\label{hyperdiffusivity}
\end{equation}
where, as in the previous Sections, we focus our attention on the case where 
$|\vec{\nabla}\psi|\ll \psi'$.
Again, if $k<k_{\sigma}$  and if  the pseudoscalar oscillations 
already took place,  
Eq. (\ref{hyperdiffusivity}) can be written as 
\begin{equation}  
\frac{\partial \vec{H}_{Y}}{\partial\tau}+ [\vec{v}\cdot\vec{\nabla}]
\vec{H}_{Y}= 
[\vec{H}_{Y}\cdot\vec{\nabla}]\vec{v},
\label{dif}
\end{equation}
where we re-expressed $\vec{\nabla} \times (\vec{v} \times \vec{H}_{Y})$
 according to the usual vector identities.
Under these approximations a fully nonlinear solution of Eqs. (\ref{dif})
and (\ref{ns}), is given by the Alfv\'en velocity \cite{bis} 
\begin{equation}
\vec{v} = \pm \frac{\vec{H}_{Y}}{\sqrt{\rho + p}}\sim  
\pm \frac{\vec{H}_{Y}}{\sqrt{\rho}}
\end{equation}
This result seems to imply that if the hypermagnetic 
field distribution is tangled
(for example as a result of the pseudoscalar oscillations), then also the 
velocity field might be tangled in the same way. 
The arguments we presented up to now hold in the case where 
$\vec{J}_{Y} \times \vec{H}_{Y} \neq 0$. We want to point out that this 
is not exactly our case. Indeed, as we showed in the previous Sections,
the maximally amplified  hypercharge modes are  
$k\sim k_{{\rm max}}$. For these modes 
$\langle \vec{H}_{Y} \cdot \vec{\nabla} \times H_{Y}\rangle$ is maximal and, 
therefore we can also argue that $\langle \vec{\nabla} \times \vec{H}_{Y} 
\times \vec{H}_{Y}\rangle \sim 0$. Thus, we can argue that our system 
is (approximately) force free \cite{cf}. Now, if the hypermagnetic field 
is not completely homogeneous we could argue from the general 
expression of the Navier-Stokes equation, that the hypermagnetic 
inhomogeneities could drive  inhomogeneities 
in the energy density according to a well known observation by Wassermann
\cite{was}. Even if the hypermagnetic field {\em would not be} force-free we 
could argue that since the electroweak epoch occurs, in our scenario, 
deep in the radiation epoch, the force term can only seed a decaying mode 
for the density fluctuations \cite{pee}.
In the case of a force free field the Navier-Stokes and diffusivity equations 
acquire a  symmetric form \cite{bis}
\begin{eqnarray}
&& \frac{\partial \vec{H}_{Y}}{\partial \tau} = \vec{\nabla}
 \times (\vec{v} \times \vec{H}_{Y}) + \frac{1}{\sigma} \nabla^2 \vec{H}_{Y}
\nonumber\\
&& \frac{\partial \vec{\omega}}{\partial \tau} = \vec{\nabla}
 \times (\vec{v} \times \vec{\omega}) + \nu \nabla^2 \vec{\omega}.
\end{eqnarray}

Let us now move to the analysis of  the kinetic equation for the 
chemical potential. We could introduce, in principle, a chemical potential
for all the perturbative strong and weak processes, Yukawa interactions of 
leptons and quarks which are in  thermal equilibrium for temperatures $T>T_c$. 
However, the processes really crucial for our purposes are the slowest 
perturbative processes related to the $U(1)_{Y}$ anomaly, namely the processes
flipping the chirality of the right electron which are in thermal equilibrium
until sufficiently late because  of the smallness of their Yukawa coupling 
\cite{relec}. Therefore the right electron number density will be diluted
thanks to perturbative processes (scattering of right electrons with the
Higgs and gauge bosons and with the top quarks because of their large Yukawa 
coupling) but it will also have a source term coming from the anomaly 
contribution.
Notice that $\Gamma$ is the (perturbative) chirality changing rate. 
Moreover, there is a second rate  appearing in Eq. (\ref{mu}) which is 
proportional to the hypermagnetic energy density. The main effect
of $\Gamma_{H}$ is to dilute the fermion number even in the 
absence of perturbative processes flipping the chirality of 
right electrons (i.e. $\Gamma=0$).

In Eq. (\ref{mu}) we see that the right electron chemical potential 
can be diluted also by diffusion. The diffusion constant $D_{R}$ appearing
in Eq. (\ref{mu}) defines the relation between the diffusion current
$\vec{J}_{R}$ and the gradient of the right electron number density
\begin{equation}
\vec{J}_{R} = - D_{R} \vec{\nabla} n_{R}.
\label{J}
\end{equation}
The right electron number dilution equation can then be written, in flat space,
as
\begin{equation}
\frac{\partial n_{R}}{\partial t} = - \frac{g'^2}{4 \pi^2} 
\vec{\cal E}_{Y}\cdot \vec{\cal H}_{Y} - \Gamma (n_{R} - n_{R}^{eq} ) -
 \vec{\nabla } \cdot\vec{J}_{R}.
\label{ndiff}
\end{equation}
Using Eqs.  (\ref{mx1})-(\ref{ohm}) together with Eq. (\ref{J}) into Eq. (\ref{ndiff})
 we can obtain Eq. (\ref{mu}). 
The right electron diffusion coefficient can be roughly estimated by taking 
into account the interactions of right electrons with hyercharge fields and it 
turns out that the typical momentum for which diffusion becomes 
effective is of the order of 
\begin{equation}
k_{D} \sim \alpha' \sqrt{\frac{T}{M_{0}}} T.
\end{equation}
For field modes $k<k_{D}$ the diffusion term can be consistently neglected
in the study of our problem.
The  hypermagnetic helicity appears to be the 
source term in the right electrons dilution equation. 
This is a quite relevant point for our considerations since 
a sizable mean magnetic helicity (possibly generated 
through psudoscalar oscillations) can influence 
the BAU present in the symmetric phase of the electroweak theory. 
In our considerations
 we will assume that the hypermagnetic diffusivity and the thermal 
diffusivity are of the same order. More precisely one can define 
the Prandtl number which usually measures, in the context of ordinary MHD, 
the relative balance between thermal and magnetic diffusivity
\begin{equation}
Pr_{m} = \nu \sigma = \frac{\sigma_0}{\alpha'}.
\label{prandtl}
\end{equation}
The 
approximation of unitary Prandtl number is often employed in 3D MHD 
simulations \cite{bis}. The 
approximation of unitary Prandtl number is often employed in 3D MHD 
simulations \cite{bis}. In this limit, hypercharge modes $k<k_{\sigma}$ 
are not damped by thermal diffusion. In our case (as in the case 
of realistic MHD simulations) $Pr_{m} >1$. 

Eq. (\ref{mu}) can be discussed in two physically different regimes.
The first one is the case where $\Gamma > \Gamma_{H}$, whereas 
the second case is the one where $\Gamma<\Gamma_{H}$. In the 
first case the rate of dilution of the chemical potential 
is dictated by the perturbative processes, whereas, in the second case,
it is determined mainly by the Abelian anomaly.

The preliminary analysis of BAU generation through Chern-Simons 
waves \cite{m3} showed that 
the case where $\Gamma> \Gamma_{H}$ may lead, potentially, to
a higher BAU and therefore we will focus our attention on the case 
where $\Gamma> \Gamma_{H}$ commenting, when needed, on the other possibility. 
Large  values of $\Gamma$ seem also consistent with the 
MSSM \footnote{In the case of the MSSM
the Yukawa couplings enhancement is certainly not the only effect
which might increase the rate. Since the particle spectrum changes
completely from the case of the MSM new diagrams will contribute to
the rate producing a further enhancement.
In particular one should perhaps assume that the right electron number
is now shared between electrons and selectrons by supergauge
interactions. Therefore it will be necessary to consider also
processes which change selectron number.} where, for large values 
of $\tan{\beta}$ (giving the ratio of the two v.e.v. of the two doublets),
the right electron Yukawa couplings can be enhanced by a factor 
$(\cos{\beta})^{-1}$. 

Since for $T>T_{c}$ the fermion number seats both in fermions 
and in hypermagnetic fields (which can carry fermionic number) 
we can expect that for  $T<T_{c}$ the fermion number sitting in the
 hypermagnetic fields should  be released in the form real fermions because 
the ordinary magnetic field (surviving after the phase transition) 
does not carry fermion number. This argument  holds
provided the EWPT
is strongly first order as it  could occur in 
the case of the MSSM also for large values of $\tan{\beta}$ \cite{mssm}. 
Notice that the presence of a hypermagnetic field might affect the dynamics 
of the phase transition itself \cite{m1,PT0}. It was recently pointed out
that, if  we have a homogeneous 
hypermagnetic field at the time of the EWPT\cite{PT2}, 
the cross-over region \cite{PT1} of the phase diagram is still 
present for sufficiently large values of the Higgs boson mass.

Thus, for $\Gamma >\Gamma_{H}$, and consequently, in the context of the MSSM,
the density of Chern-Simons number present before the phase transition can be
 released in fermions which will have some chance of 
not being destroyed by sphalerons processes \cite{m1} and then the 
density of baryonic number can be related to the integrated anomaly in the 
usual way \cite{m2,m3}.
 The BAU can then be expressed 
directly in terms of the hypermagnetic helicity as
\begin{equation}
\frac{n_{B}}{s}(\vec{x}, t_{c}) =
\frac{\alpha'}{2\pi\sigma_c}\frac{n_f}{s}
\frac{{\vec{{\cal H}}}_{Y}\cdot \vec{\nabla}\times
{\vec{{\cal H}}}_{Y}}{ \Gamma + \Gamma_{{\cal H}}}\frac{\Gamma M_{0}}{T^2_c},
\label{BAU}
\end{equation}
[$s = (2/45) \pi^2 N_{eff}T^3$ is the entropy density and $N_{eff}$ is the
 effective number of massless degrees of freedom ($106.75$ in the minimal 
standard model)].
In the minimal standard model (MSM) $\Gamma \sim T (T_{R}/M_{0}) $ where 
$T_{R}\sim 80 $ TeV.  If
 $\Gamma > \Gamma_{H}$  the hypermagnetic field has to be sufficiently 
weak and/or $\Gamma$ is quite large. For example in the minimal supersymmetric
standard model (MSSM) the right electron Yukawa coupling is larger than in 
the MSM case so that $T_{R}$ can be larger by a factor of the order of $10^{3}$
for value of $\tan{\beta}\sim 50$. If the hypermagnetic 
background is sufficiently intense, then, $\Gamma_{H} >\Gamma$. 
This last case can arise in the MSM. If we suppose that 
$\Gamma_H >\Gamma$ we run, however, into a contradiction
among our assumptions. In fact if we are just in the context of the MSM
the phase transition cannot be strongly first 
order for Higgs boson masses larger than the $W$ boson mass. This result
is not crucially modified by the presence of a strong hypermagnetic
background \cite{PT2,PT1}. Therefore, if $\Gamma_{H}>\Gamma$ the 
BAU possibly generated for $T>T_{c}$ by our mechanism will be washed out 
by a subsequent stage of thermal equilibrium. However, even if we would 
assume that, thanks to some presently unknown phenomena, the EWPT would be 
strongly first order in the MSM, we will be able to show in the next section 
 that the BAU obtained in the case $\Gamma_{H}>\Gamma$ is too small to be 
phenomenologically relevant.

In the case where $\Gamma>\Gamma_{H}$ we have that 
the BAU can be expressed as 
\begin{equation}
\frac{n_{B}}{s}(\vec{x}, t_{c}) = \frac{\alpha'}{2\pi \sigma_c} 
\frac{n_{f}}{s}\langle {\vec{{\cal H}}}_{Y}\cdot \vec{\nabla}\times
{\vec{{\cal H}}}_{Y}\rangle \frac{M_0}{T_{c}^2}.
\label{BAU1}
\end{equation}
There are two (physically complementary) 
situations 
where Eqs. (\ref{BAU})--(\ref{BAU1}) can be applied. The first one is the case of a 
stochastic hypermagnetic background. In this case $\langle n_{B}/s \rangle$ is 
strictly zero but its fluctuations (i.e. $\langle(n_{B}/s)^2\rangle$) 
are non vanishing \cite{m3}. On the other hand in our present case the 
hypermagnetic field distribution is not topologically trivial because of the
action of the pseudoscalar oscillations and, therefore, 
$\langle n_{B}/s \rangle \neq 0$.

\renewcommand{\theequation}{7.\arabic{equation}}
\setcounter{equation}{0}
\section{Phenomenological Considerations}

As a preliminary 
exercise let us consider the case of a Chern-Simons wave configuration  
\cite{hyp,m1} (see also \cite{m3}). 
Suppose that at some moment of time the universe is populated 
by parity non-invariant configurations of the type discussed in Eq. 
(\ref{conf1}). using the hypermagnetic field configuration obtainable 
from eq. (\ref{conf1}) into  Eq. (\ref{BAU}) we get that, in the 
limit $\Gamma>\Gamma_{H}$ 
(we work, for simplicity, in flat space), the BAU can be written as
\begin{equation}
\left(\frac{n_{B}}{s}\right)(\vec{x}, t_{c}) \simeq  
\frac{\alpha'}{2~\pi~\sigma_{c}} \left(\frac{n_{f}}{s}\right) 
\left(\frac{k_{0}}{T_{c}}\right) \left(\frac{M_{0}}{T_{c}}\right)
{\cal H}^2(t_{c})\simeq 
10^{10} \left(\frac{k_{0}}{T_{c}}\right) 
\left(\frac{ {\cal H}^2}{T^4_{c}}\right),
\end{equation}
As we can see, a substantial BAU can be achieved  provided
the primordial value of the initial hypermagnetic field is large with 
respect to vacuum fluctuations. Suppose in fact that ${\cal H}^2$ is
only built up from vacuum fluctuations. Then ${\cal H}^2\sim k_0^4$.
Taking now into account that, at the electroweak scale,
the conductivity momentum is, roughly, 
$k_{\sigma} \sim 10^{-7}~T_{c}$
we have that the  obtained BAU for the highest possible 
mode which could survive in the plasma is of the order of
$10^{-25}$. Too small to be relevant. In the framework 
of this  example we see that there exist always
a minimal hypermagnetic energy density which is compatible 
with the BAU. The reason for this fact is the existence
of a {\em finite} diffusivity scale. In our case
 the smallest magnetic energy density we can
consider is given by ${\cal H}^2 \sim 10^{-14} T_{c}^4$. For 
smallest values the BAU cannot be reproduced not even 
taking the maximal frequency of the spectrum, namely
$k_0\sim k_{\sigma}$. Different Chern-Simons condensates will
have, of course, different numerical features and
in  the following we will try to understand if the 
presence of dynamical pseudoscalar can give a dynamical 
justification of these preliminary considerations.

We will address, separately, the 
case of amplification occurring during inflationary phase 
and during radiation dominated phase. 
If the pseudoscalar oscillations take place in an inflationary 
phase, the production of Chern-Simons condensates from the
vacuum fluctuations of the associated hypermagnetic fields 
is possible. If, on the other hand, the oscillations will occur in the 
radiation dominated epoch there is no chance of getting 
interesting phenomenological values starting 
from vacuum fluctuations. The reason for this 
behavior is very simple: if the pseudoscalar
oscillations arise during an inflationary epoch the 
amplification of the hypermagnetic fields is not
damped by the finite value of the conductivity. 
\begin{figure}
\begin{center}
\begin{tabular}{|c|c|}
      \hline
      \hbox{\epsfxsize = 7 cm  \epsffile{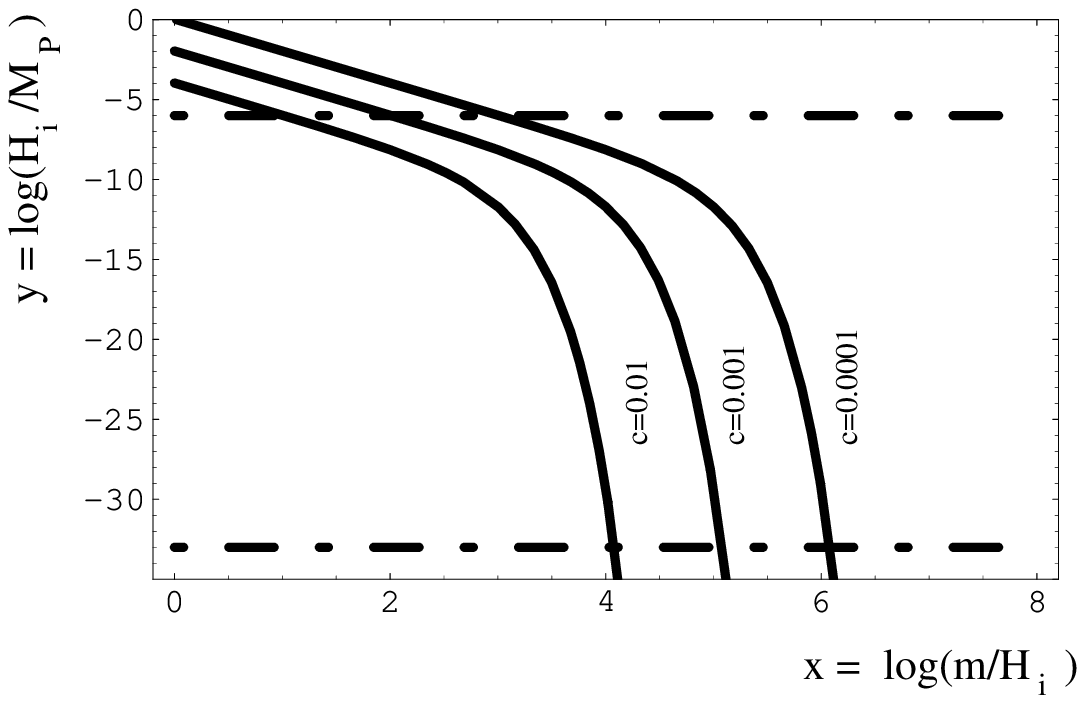}} &
      \hbox{\epsfxsize = 7 cm  \epsffile{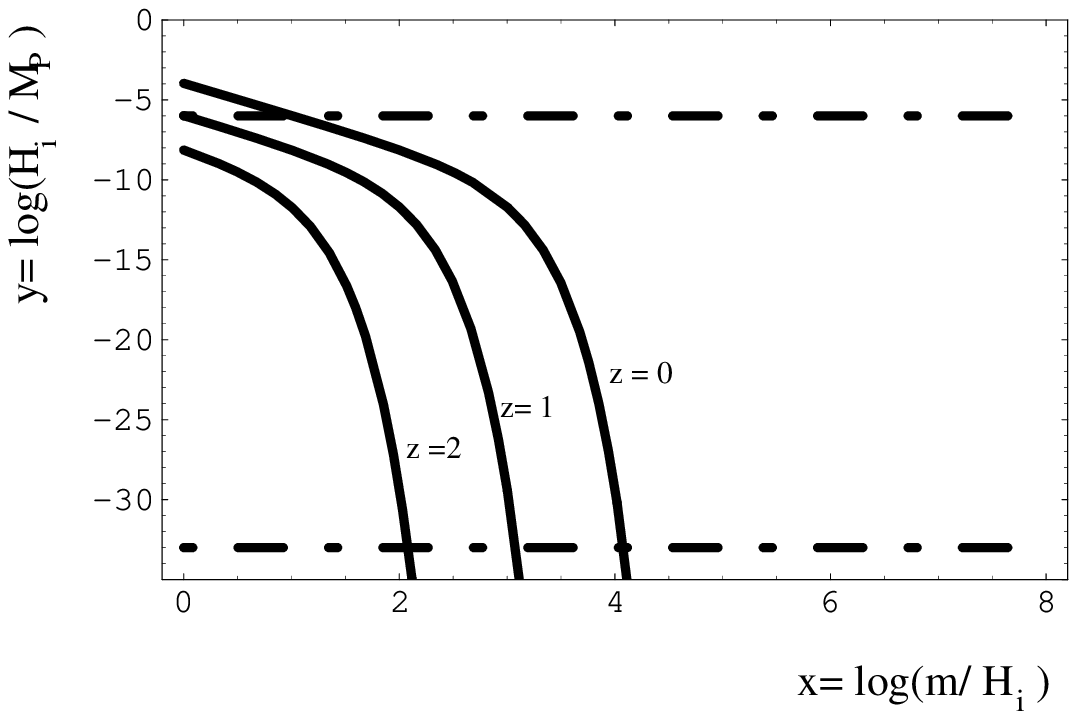}} \\
      \hline
\end{tabular}
\end{center}
\caption[a]{We illustrate the result obtained in Eq. (\ref{BAU3}). 
The two dot-dashed lines in the plots 
 correspond to $H_{i} \sim 10^{-6} M_{P}$  
and $H_{i} \sim 10^{-33}M_{P}$. The physical region of our parameter space 
is within the two dot-dashed straight lines 
since the maximal (inflationary) 
curvature scale cannot exceed $10^{-6}M_{P}$ (as required by 
the measurements
 of the CMB anisotropy) and cannot also go below $10^{-33}M_{P}$ (where the 
EWPT takes place). Moreover, we would expect the 
maximal inflationary curvature  scale to be larger than the
 curvature scale of the electroweak epoch since we want, by that time,
 the Universe to  be dominated by radiation. In the left picture we report,
 in full lines, various curves leading to a BAU of the order of $10^{-10}$ for 
different values of the coupling constant of the hypercharge field to the
 pseudoscalar. In the right picture we fixed $c=0.01$ but we left
 $z= \log_{10}{(\psi_0/M)}$ free to change from $0$ to $2$.In order to 
have a reasonable BAU we have to stay within the dashed lines but also 
above the thick lines. We can clearly see that this can be achieved 
not only in the case of large $x$ and small $y$, but also for $x\sim 1$,$2$ 
and $y\sim -6$,$-10$ provided either the coupling constant is not too small 
and/or the initial amplitude of oscillation is larger than $M$ (i.e. $z>1$). 
In this and in the following two figures we always took $N_{eff}=106.75$ 
and $\sigma_0 =70$
since we assumed our hypermagnetic modes to  re-enter in the 
symmetric phase of the standard electroweak theory. }
\label{f1}
\end{figure}

\subsection{BAU from Hypermagnetic  Knots  Generated at the end of Inflation}

Eqs. (\ref{BAU})--(\ref{BAU1})  can be 
 explicitly evaluated using the form 
of our Bogoliubov coefficients given in 
Eqs. (\ref{c1})-(\ref{c2}) with the result that 
for $\omega\sim \omega_{{\rm max}}$ the mean value of the 
hypermagnetic helicity is 
\begin{equation}
\langle \vec{H}_{Y} \cdot \vec{\nabla} \times \vec{H}_{Y} \rangle 
= \frac{\omega_{{\rm max}}^5}{\pi^2} e^{ c (\frac{m}{H_i}) (\frac{\psi_0}{M}) }
 e^{ -2 (\frac{\omega_{{\rm max}}}{\omega_{\sigma}})^2}.
\label{BAU2}
\end{equation}
Inserting Eq. (\ref{BAU2}) into Eq. (\ref{BAU1}) and  taking into account
 that the maximally amplified
 frequency  $\omega(t_i) \sim \omega_{{\rm max}}(t_i)
 = m(c/2)(\psi_0/M)$, red-shifted at the electroweak epoch, reads
\begin{equation}
\omega_{{\rm max}}(t_{\rm ew}) = \frac{c}{2} \biggl(\frac{\psi_0}{M}\biggr) 
\biggl(\frac{m}{H_i}\biggr)\sqrt{\frac{H_i}{M_{P}}} (N_{eff})^{1/4}T_{c},
\label{freq}
\end{equation}
we can obtain the explicit form of the logarithmic spectrum of the BAU, namely 
\begin{equation}
\frac{d (n_{B}/s)}{d\log{\omega}} = \frac{g'^2 c^5}{512} \frac{45 n_{f} 
(N_{eff})^{1/4}}{\sigma_0 \pi^6}\frac{M_0}{T_{c}} 
\biggl(\frac{\psi_0}{M}\biggr)^5
\biggl(\frac{m}{H_{i}}\biggr)^5 \biggl( \frac{H_{i}}{M_{P}}
\biggr)^{\frac{5}{2}} e^{c (\frac{m}{H_{i}}) (\frac{\psi_{0}}{M})}
e^{-2(\frac{\omega_{{\rm max}}}{\omega_{\sigma}})^2}.
\end{equation}

\begin{figure}
\begin{center}
 \begin{tabular}{|c|c|}
      \hline
      \hbox{\epsfxsize = 7 cm  \epsffile{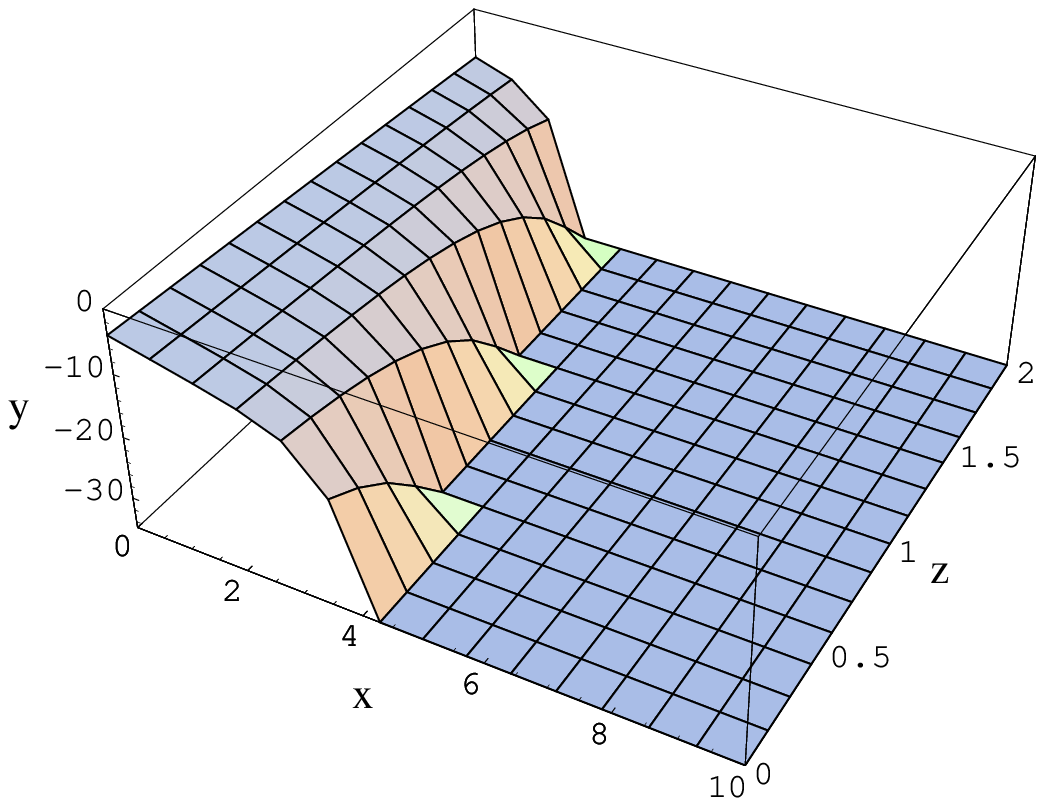}} &
      \hbox{\epsfxsize = 7 cm  \epsffile{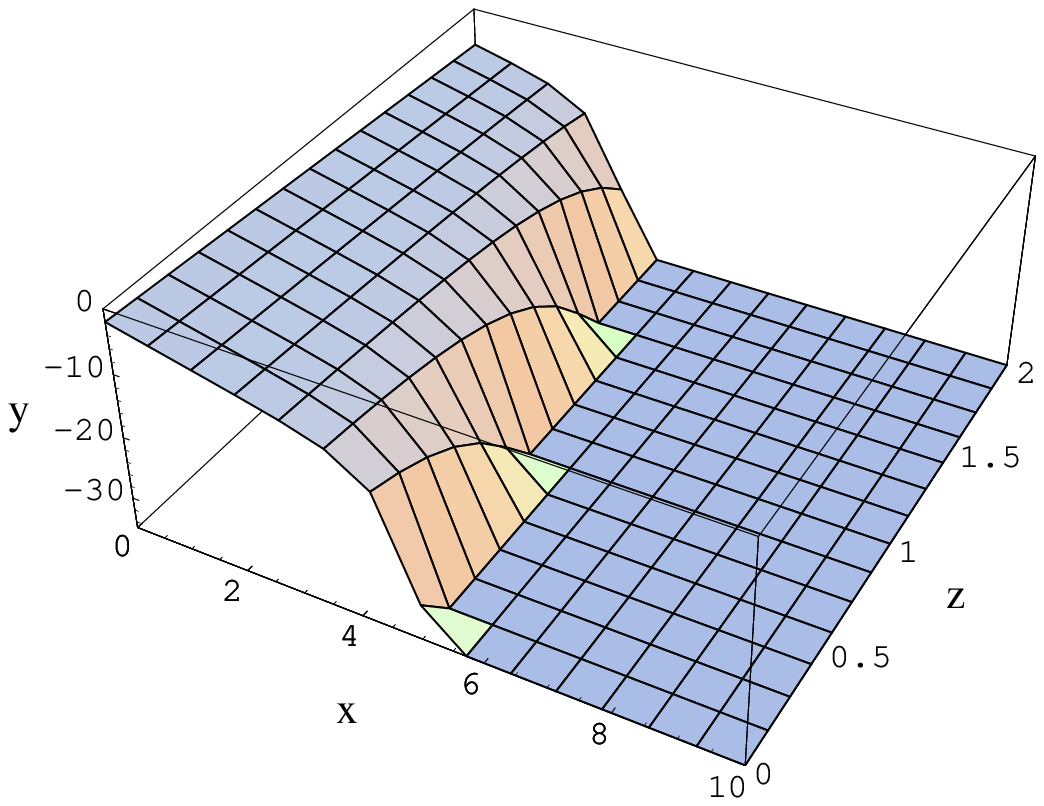}} \\
      \hline
\end{tabular}
\end{center}
\caption[a]{
At the left we illustrate a surface of constant BAU for a coupling constant
of the order of $c\sim 0.01$ as a function of $x$, $y$ and $z$. At the 
right we plot the same quantity [i.e. Eq. (\ref{BAU3})] but for $c=0.001$.}
\label{f2}
\end{figure}

We want now to compare Eq. (\ref{BAU2}) with the 
required value of the baryon asymmetry. Defining $x= \log_{10}{( m/H_{i})}$,
$y= \log_{10}{(H_{i}/M_{P})}$ and $z=\log_{10}{(\psi_0/M)}$ we have that, 
if we want $n_{B}/s \gaq 10^{-10}$, we have to demand, from 
Eqs. (\ref{BAU})--(\ref{BAU1}), that 
\begin{equation}
y\geq - 8.5 -\frac{1}{10}\log_{10}{N_{eff}} + \frac{2}{5}\log_{10}{\sigma_0}
- 2 \log_{10}{c} - 2 (x + z) 
- \frac{ 2~c}{5} 10^{(x + z)} \log_{10}{e},
\label{BAU3}
\end{equation}
where $\sigma_0 = \sigma_{c}/T\sim 70$--$100$ \cite{sigma0}.
Eq. (\ref{BAU3}) illustrated in Fig. \ref{f1} and \ref{f2}. In the thick 
lines we represent curves of constant BAU (of the order of $10^{-10}$) as function
of the two parameters of the model, namely the maximal inflationary scale and 
the mass of the pseudoscalar. We can see that for inflationary
 curvature scales $H_{i}\sim 10^{-6}$--$10^{-10}~M_{P}$ 
the mass has to be of the order of $10^2~ H_{i}$. At the same time,
 provided the parameter space 
does not hit the electroweak curvature scale 
(lower dot-dashed line in Fig \ref{f1}),
the inflationary scale can be reduced and then, for 
$H_{i} \sim 10^{-15}$--$10^{-20}~M_{P}$, we 
should require masses of the order 
of $10^{3}$--$10^{4}~H_{i}$ in order to achieve the required value of the BAU.
If, initially, the field is not
 in its vacuum state the hypermagnetic helicity is given by
\begin{equation}
\langle \vec{H}_{Y} \cdot \vec{\nabla} \times \vec{H}_{Y} \rangle =
\frac{\omega^5_{\max}}{\pi^2} |{\cal A}(\omega_{{\rm max}})|^2 
e^{ c (\frac{m}{H_{i}}) (\frac{\psi_0}{M})}
\label{BAU4}
\end{equation}
Inserting Eq. (\ref{BAU4})  into Eq. (\ref{BAU1}) 
the logarithmic BAU spectrum will depend explicitly on
 $r(\omega_{{\rm max}})$ which is the critical fraction of 
energy density stored in the initial (topologically trivial)
hypermagnetic field distribution.

\begin{figure}
\centerline{\epsfxsize = 7 cm  \epsffile{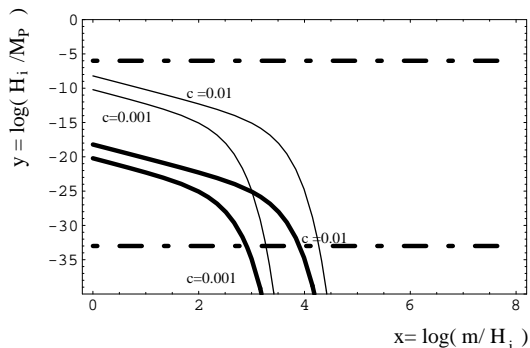}} 
\caption[a]{We illustrate Eq. (\ref{BAU5}) in the case of $z=0$ and 
$r(\omega_{{\rm max}})\sim 10^{-10}$. The acceptable 
region of parameter space  lies, as usual,
 between the two dot-dashed lines and  above 
the full (thick) lines. 
If the initial state of the hypercharge fields is the vacuum the 
maximal inflationary scale compatible with a BAU  of the order of $10^{-10}$ 
gets lower  if compared with the case reported in 
Fig. \ref{f1}. We also illustrate (full thin lines) 
the same case reported in the 
thick lines but for a lower value of 
$r(\omega_{{\rm max}})$, namely, $10^{-15}$.}
\label{f3}
\end{figure}

In order to be compatible with value of the BAU obtained in the context of the
homogeneous and isotropic BBN scenario we have 
to require
\begin{equation}
y \gaq - 46.88 -\frac{1}{2}\log_{10}{N_{eff}}-
2 \log_{10}{\sigma_0}- 2 \log_{10}{r(\omega_{{\rm max}})} 
- 2 \log_{10}{c} -2 (x + z) - 2 ~c ~10^{x + z}~ \log_{10}{e}.
\label{BAU5}
\end{equation}
By looking at Eqs. (\ref{BAU2})--(\ref{BAU3}) together 
with  Eq. (\ref{BAU5}) we see that in order 
to achieve the observed value of the BAU with reasonable values of the
pseudoscalar  mass in curvature units (i.e. $m/H_{i}\sim 10^2$) we have to 
postulate an inflationary phase at curvature much higher than the electroweak
 epoch $H_{i} \sim 10^{-10}M_{P}$. If, initially, the hypercharge
 modes are not in the vacuum  we can 
see that if $m\gaq H_{i}$ and $H_{i} \sim 10^{-20} M_{P}$the observed value of 
the BAU can be achieved provided $r(\omega_{{\rm max}}) \sim 10^{-10}$. 
This aspect is illustrated in Fig. \ref{f3}. 
By lowering the energy density of the 
primordial stochastic background we are pushed towards higher
 inflationary scales
and/or  higher pseudoscalar masses.
 
Therefore, the generation of hypermagnetic helicity is quite
likely if the pseudoscalar oscillations take place during an inflationary phase
for a wide range of masses. If we want the BAU generation to occur at very 
high curvature scales (i.e. $H_{i}\sim 10^{-6}$--$10^{-10}~M_{P}$) we are also
 led towards  high mass scales (i.e. $m\gaq 10^{2} H_{i}
\sim 10^{8}$--$10^{12}$ TeV). 
The exponential amplification appearing in the Bogoliubov 
coefficients is responsible of the sharp cut in the plots reported in Figs.
\ref{f1}-\ref{f4}. Thus, if we take the curvature scale to be sufficiently
low (say, for instance, $10^{-21}M_{P}$),  then we can get a sensible BAU 
for masses $m\gaq 10^{4} H_{i} $, i.e. masses of the order of few hundred
GeV at the time the amplification occurred. Notice that the inclusion 
of stimulated emission goes exactly in this direction, namely it allows,
more easily, low curvature scales.

We finally notice that, from Eq. 
(\ref{freq}), the 
the maximal amplified frequency (red-shifted at the electroweak 
scale ) is always smaller than the diffusivity frequency 
at the electroweak scale
(i.e. $\omega_{\sigma}(t_{\rm ew}) \sim 10^{-7}~T_{\rm ew}$). This
 guarantees that  the 
obtained BAU will not be washed out by 
hypermagnetic diffusion.

\subsection{BAU from Hypermagnetic Knots Generated During 
a Radiation Epoch}

We are now going to consider the phenomenological implications of 
pseudoscalar oscillations occurring in a radiation dominated phase.
In order to get a  more explicit expression of the BAU in this case
let us insert Eq. (\ref{con3}) into Eq. (\ref{BAU}). Since  
$\omega_{{\rm max}}/\omega_{\sigma}<1$ we  have that the BAU can be written as
\begin{equation}
\frac{n_{B}}{s} \simeq \frac{45 n_f }{8\pi^3 \sigma_0} ~c~ \alpha'
 \frac{\Delta\psi}{M} 
r(\omega_{{\rm max}}).
\label{BAUrad}
\end{equation}
It is interesting to notice that, as far as the maximal frequency is concerned,
the BAU does not depend upon the pseudoscalar mass but only 
upon the pseudoscalar coupling constant.
Let us estimate, very roughly, the size of the BAU emerging 
from Eq. (\ref{BAUrad}).
 Suppose that $N_{eff} =106.75$ 
(same particle content of the standard model) and that $c =0.01$. Thus 
for $\sigma_0\sim 70$ we get that $n_b/s \sim 10^{-7} r(\omega_{{\rm max}})$.
Thus the BAU can be reproduced if $ r(\omega_{{\rm max}}) \sim 10^{-3}$.
Therefore, the presence of an initial background
of hypermagnetic fields is essential
 as in the case of the Chern-Simons 
wave discussed at the beginning of the present Section. 
Suppose in fact that 
that ${\cal A}(\omega)=1$. In this case the initial stochastic background 
is  formed by vacuum fluctuations. Therefore 
$r(\omega_{{\rm max}}) = N^{-1}_{eff}
 (\omega_{{\rm max}}/T)^4 \sim 10^{-33}$ giving 
a BAU of the order of $10^{-40}$. 

\begin{figure}
      \centerline{\epsfxsize = 8 cm  \epsffile{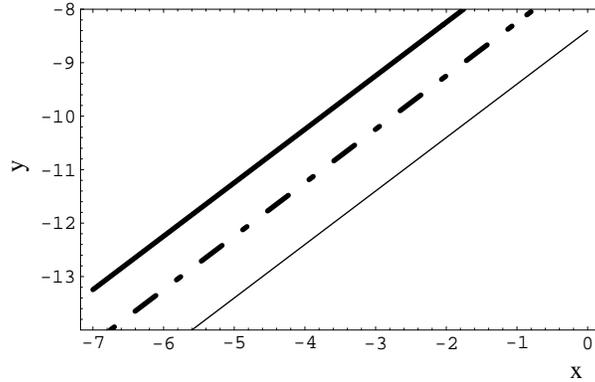}} 
\caption[a]{We illustrate the logarithmic variation of the BAU computed 
in Eq. (\ref{BAUrad2}) for different sets of parameters. More precisely
we have $c=0.01$, $\sigma_0 =70$ (full thick line), $c=10^{-3}$, 
$\sigma_0=70$ (dot-dashed line), $c=10^{-4}$, $\sigma_0=100$ (full thin line).}
\label{f4}
\end{figure}

The logarithmic variation of the BAU 
as a function of the parameters of the model can be expressed as 
\begin{equation}
y = - 2.4 + \log_{10}{c} - \log_{10}{\sigma_0} 
+ \log_{10}(\frac{\Delta\psi}{M}) + x
\label{BAUrad2}
\end{equation}
where now $x = \log_{10}{r(\omega_{{\rm max}})}$ and $y= \log_{10}{(n_{B}/s)}$.
From this expression illustrated in Fig. \ref{f4} we can draw two
conclusions. The first one is that the BAU possibly
 generated  during the radiation dominated epoch depends upon the 
pseudoscalar coupling but not upon the pseudoscalar mass. This
happens  because we estimated the BAU arising in the case 
of the maximally amplified frequency whose 
actual value is dictated by Eq. (\ref{exp}).
 The smaller is the coupling the larger has to be the 
primordial hypermagnetic background. 
The second observation is that the possible 
explanation of the BAU through Eqs. (\ref{BAUrad})--(\ref{BAUrad2}) depends 
crucially upon the strength of the EWPT. In fact, in
order to derive Eq. (\ref{BAUrad}) we demanded  the rate of right
electron chirality flip processes (i.e. $\Gamma$) to be larger than 
$\Gamma_{H}$. As we discussed,
 this hypothesis is certainly  not forbidden in the case of the MSSM. 
One could also wonder  if the MSM physics would be enough
to implement our scenario. Let us assume, for a moment,
to be in the framework of the MSM. Then, as argued in \cite{m2,m3}, 
$\Gamma_{H}>\Gamma$. The principal problem with this 
assumption is the strength of the EWPT.
 In fact we know that even in the presence of a strong magnetic field
 \cite{PT2} the typical cross-over behavior \cite{PT1} of the phase transition
 does not change \footnote{More precisely, for sufficiently small hypermagnetic
backgrounds (i.e. $|\vec{H}_{Y}|/T_{c}^2 \laq 0.3$) 
the phase transition is stronger but it turns 
into a cross-over for sufficiently large Higgs masses. It would 
be interesting to understand if for sufficiently strong hypermagnetic 
backgrounds the Ambjorn-Olesen phase can be generated \cite{AO}}. 

However, even if the phase transition would be 
strongly first order our mechanism  would be ineffective. 
Indeed, let us compute the BAU in the case where 
$\Gamma_{H} >\Gamma$. We have that the BAU can be expressed 
as \cite{m2}
\begin{equation}
\frac{n_{B}}{s} = \frac{\alpha'}{2 \pi\sigma_c} \frac{n_f}{s} 
\langle \frac{\vec{H}_{Y}\cdot\vec{\nabla}\times\vec{H}_{Y}}{\Gamma_{H}}
\rangle 
\frac{\Gamma M_0}{T_{c}^2}.
\end{equation}
The expectation value appearing in the previous equation involves 
a highly non local operator which can be evaluated for scales 
larger than the diffusivity scale with the result that \cite{m3}
\begin{equation}
\langle \frac{\vec{H}_{Y}\cdot\vec{\nabla}\times\vec{H}_{Y}}
{|\vec{H}_{Y}|^2}\rangle
\sim \frac{\langle \vec{H}_{Y}\cdot\vec{\nabla}\times\vec{H}_{Y}\rangle  }{ 
\langle |\vec{H}_{Y}|^2 \rangle }.
\end{equation} 
Thus,  we get that the BAU can be expressed as 
\begin{equation}
\frac{n_{B}}{s} \simeq 0.04~c~ \biggl(\frac{\Delta\psi}{M}\biggr)
 \biggl(\frac{T_{R}}{M_0}\biggr).
\end{equation}
where $T_{R}\sim 80$ TeV is the right-electron equilibration temperature 
\cite{relec}. For typical values of the parameters 
(the same we used in the case $\Gamma>\Gamma_{H}$ the BAU is of the order of 
$10^{-18}$. We would be tempted to say that for large $T_{R}$ the BAU
 could be recovered. Unfortunately this assumption would contradict 
our initial hypotheses since it would mean that we  go beyond
the MSM and that, consequently, we would have  $\Gamma_{H}>\Gamma$. 
Notice that, as previously discussed \cite{mmm}, the results of our analysis 
are not in agreement with the statement that the BAU can be generated 
from dynamical pseudoscalars in the MSM, for hypermagnetic modes 
of the order of $k_{{\rm max}}\sim T$ \cite{ob} and independently 
of the specific value of the perturbative rate of right electron 
chirality flip processes.

We want to observe,  finally,  that 
some mechanisms  (like the breaking of conformal invariance) 
operating at some higher energy scale \cite{ggv} can 
generate a {\em topologically trivial} hypermagnetic background whose 
energy density is small in critical units but anyway quite large if 
compared to the energy density of the vacuum fluctuations. 
Since we are at high 
conductivity, the dissipation of the seeds is protected from being washed out
by the hypermagnetic flux conservation. Then, at some temperature
 (in the TeV range) pseudoscalar oscillations will begin. 
At finite conductivity the pseudoscalar oscillations do not
crucially amplify the initial hypermagnetic background but they twist its 
topology leading to a non vanishing Chern-Simons number density and, 
ultimately, to the BAU. 

\renewcommand{\theequation}{8.\arabic{equation}}
\setcounter{equation}{0}
\section{Parity Breaking from Axions?}

The same mechanism studied in the case of HK can be indeed 
exploited in order to produce ordinary {\em magnetic} knots for temperatures 
smaller than $100$ GeV. The natural candidate would then be, in this case,
the ordinary QCD axion. We will start by recalling the possible implications
of global parity breaking in ordinary MHD and we  will then show that 
the amount of global parity breaking induced by axion oscillations is 
indeed quite minute.

\subsection{Global Parity Breaking in MHD}
If, at some scale \footnote{We denote with $\vec{H}$ the {\em ordinary } magnetic field
as opposed to the hypermagnetic field.}
$\langle\vec{H}\cdot\vec{\nabla} \times \vec{H} \rangle \neq 0$, than we also
 have that the Alfv\'en (stationary) flow will be tangled in the same 
way the related magnetic field is tangled since 
$\langle\vec{v}\cdot\vec{\omega}\rangle 
= \langle \vec{v}\cdot\vec{\nabla}\times\vec{v}\rangle\neq 0$. This 
last condition together with the incompressible closure 
$\vec{\nabla}\cdot \vec{v}=0$ implies  that also the  velocity field is 
formed by a collection of closed loops whose degree of 
knottedness can be estimated from $\vec{v}\cdot\vec{\omega}$. Notice that 
in the limit of negligible thermal and magnetic dffusivity both the magnetic 
and the velocity field can be treated with topological techniques 
\cite{mof,ber}. When the thermal and magnetic diffusivity are adiabatically
switched on (or, more physically, when we approach the magnetic and thermal
diffusivity scales) various dissipative phenomena arise and turbulence sets in.
We do not want even to attempt to address the very rich structure arising in
 MHD turbulence. However, we simply want to notice that there are
crucial differences if the turbulence arises starting from a parity 
invariant or from a parity non invariant configuration of the 
bulk velocity field 
\cite{parker,rev3}. It has been actually  observed long ago that 
\cite{vainshtein}, apparently, cosmic turbulence occurs usually 
in rotating bodies with density gradients. This results in a violation of the
mirror symmetry of the plasma since it is possible, by taking the scalar 
product of the vorticity with the density gradient, to construct a 
pseudoscalar. The relevance of (global) parity breaking in MHD can be also 
appreciated by looking at the turbulent dynamo mechanism 
\cite{rev3,vainshtein}. The idea is that in astrophysical and cosmological 
plasmas the correlation scale of the velocity field is normally
shorter than the correlation scale of the magnetic field. Under this assumption
from Eq. (\ref{dif}) it possible to obtain a dynamo equation by carefully 
averaging over the velocity field over scales and times exceeding 
the characteristic correlation scale and time $\tau_{0}$ of the velocity field.
In this approximation the magnetic diffusivity equation can be written as
\begin{equation}
\frac{\partial\vec{H}}{\partial\tau} =
\alpha(\vec{\nabla}\times\vec{H}) +\frac{1}{\sigma}\nabla^2\vec{H}, ~~~
\alpha 
= -\frac{\tau_{0}}{3}\langle\vec{v}\cdot\vec{\nabla}\times\vec{v}\rangle,
\label{dynamo}
\end{equation}
where $\alpha$ is the so-called dynamo term. In Eq. (\ref{dynamo})
 $\vec{H}$ is
the magnetic field averaged 
over times larger than $\tau_{0}$, which is the typical correlation
time of the velocity field). We can clearly see that the crucial
requirement for the all averaging procedure we described is that the
turbulent velocity field has to be ``globally'' non-mirror-symmetric.
If the velocity field {\em is}
parity invariant (i.e. no vorticity for scales comparable with the
correlation length of the magnetic field), then the dynamics of the
infrared modes is decoupled from the velocity field since, over those
scales, $\alpha =0$. Therefore, for our purposes the $\alpha$ term will
be a good measure of the degree of knottedness associated with the velocity
field.

Needless to say that that there is a puzzle associated with the turbulent 
dynamo mechanism \cite{rev3}. On one hand we {\em see} that our galaxy has 
an $\alpha$ term which can be roughly estimated to be $\alpha_{G} \sim 0.8 
\times 10^{5}$ cm/sec in the interval $3~{\rm Kpc}\laq r\laq 12~{\rm Kpc}$ (it
is assumed that $|\vec{\omega}| \sim 10^{-15} ~ {\rm sec}^{-1}$) 
\cite{rev3}. On the
other hand we have to postulate that {\em some} alpha term had to be present 
also at the time of the protogalactic collapse  
(occurred around $t_{gc} \sim 10^{15} $ sec). This is because we want the 
dynamo term to amplify, by differential rotation the primordial magnetic seeds
up to the observed value of themagnetic field observed in spiral galaxies.
Thus, the parity breaking of the galactic turbulence seems to be more a 
consequence of some primordial (global) parity breaking than a theory of the
 galactic dynamo action.

If one wants to postulate some global parity breaking in the plasma the 
obvious question which arises is when this phenomenon could occur in the 
early Universe. It should certainly be useful for the dynamo action to have 
some sort of global parity breaking at the decoupling epoch 
($t_{dec} \sim 10^{11} sec$) which is not so far (for cosmological standards)
from the time of the protogalactic collapse. In principle, the parity
breaking could occur even before but the only problem could be, in this case,
that many of the known phenomena occurring in the early Universe are normally
discussed in terms of a mirror-symmetric plasma. For example the 
present BBN calculations do not include, to the best of our knowledge, 
possible effects of global parity violation \cite{revN,IBBN} and we would guess
that to have global parity breaking might be a problem for nucleosynthesis.

The axion oscillations starting at a temperature of the order of $1$ GeV can 
certainly produce a source of global parity breaking. The question is how 
large is the induced $\alpha$ term in the primordial plasma. Our aim is 
now to estimate  the order of magnitude of the parity breaking induced 
through primordial electromagnetic helicity.

\subsection{Parity Breaking induced by ordinary Magnetic Knots }

As it is well known in axion models \cite{kim,PQ}  an extra (global) 
$U_{PQ}(1)$ symmetry complements the MSM. This symmetry is broken at the 
Peccei-Quinn scale $F_{a}$ and leads to a  dynamical pseudo Goldstone boson
(the axion) which acquires a small mass because of soft instanton effects 
at the QCD phase transition. If an axionic density is present in the early 
Universe , bounds can be obtained for the Peccei-Quinn symmetry breaking scale.
These bounds together with astrophysical bounds (coming mostly from 
red-giants cooling and from SN 1987a) leave a window opportunity 
$10^{10} {\rm GeV} < F_{a} < 10^{12} {\rm GeV}$ \cite{raffelt}.

There are different realizations of axion models. In some models 
\cite{kim,svz} quarks and leptons do not carry the Peccei-Quinn charge,
 whereas in other realizations the usual standard model leptons are not 
singlet under the Peccei-Quinn symmetry \cite{kim,DFSZ}. We will not enter
into the details of these two models since what matters for our consideration
is, at least in the very first approximation,
 how the axion couples to  the anomaly.

There are also different cosmological realizations of axionic models. If an 
inflationary epoch takes place prior to the usual radiation dominated regime,
 the Peccei-Quinn phase can be aligned by today \cite{linde}. 
In the absence of an inflationary phase it is likely that as the Universe 
passes through the Peccei-Quinn phase transition a network of cosmic 
strings will be generated  via the Kibble mechanism \cite{ve}. We will 
assume that an inflationary phase took place well before
 ($H\leq 10^{-6} M_{P}$) the onset of the EWPT.

If the temperature of the Universe exceeds the critical temperature of the
EWPT the effective action describing the 
interaction of axions with the Standard model gauge fields can be written as 
\begin{equation}
S_{eff} = \int d^{4} x \sqrt{-g}\biggl[ - \frac{\psi}{F_{a}} \biggl( 
\frac{g_{s}^2}{32 \pi^2} G^{a}_{\mu\nu} \tilde{G}^{\mu\nu a} + c_{\psi W} 
\frac{ g^2}{32 \pi^2}  W^{a}_{\mu\nu} \tilde{W}^{\mu\nu a} +  c_{\psi Y} 
\frac{g'^2}{32 \pi^2} Y_{\mu\nu}\tilde{Y}^{\mu\nu} \biggl)\biggr]
\label{axionaction1}
\end{equation}
where $g_{s}$, $g$ and $g'$ are, respectively, the $SU_{c}(3)$, $SU_{L}(2)$
 and $U_{Y}(1)$ coupling constants ($c_{\psi W}$ and $c_{\psi Y}$ are 
numerical constants of order 1 which depend upon the specific axion model).

For $T\leq T_{c}$ the effective action becomes
\begin{equation}
S_{eff} = \int d^{4} x \sqrt{-g}\biggl[ - \frac{\psi}{F_{a}} \biggl( 
\frac{g_{s}^2}{32 \pi^2} G^{a}_{\mu\nu} \tilde{G}^{\mu\nu a} +  
c_{\psi\gamma} \frac{e^2}{32 \pi^2} F_{\mu\nu}\tilde{F}^{\mu\nu} \biggl)\biggr]
\label{axionaction2}
\end{equation}
where $F_{\mu\nu}$ is the ordinary (electromagnetic) gauge field strength.

There are important differences between the axion evolution for 
temperatures larger and smaller than the EWPT 
temperature. 
In fact \cite{mms}, for $T\geq T_{c}$ the axion evolution equation is not 
only modified by the Hubble damping factor, but also by damping induced by
 the  QCD sphalerons which are not suppressed at high temperatures. 
This effect leads to an equation for 
$\psi$ which differs from the one introduced in the previous section and
 which can be written as 
\begin{equation}
\ddot{\psi} + (3 H + \gamma) \dot{\psi} =0
\end{equation}
where we dropped for simplicity the potential term. It turns out that
\cite{mms}
\begin{equation}
\gamma = \frac{\Gamma_{{\rm sph}}}{F_{a}^2 T} \simeq 
\frac{\alpha_{s}^4 T^3}{F_{a}^2}
\end{equation}
(where $\alpha_{s} = g_{s}^2/4\pi$). Thus, if $F_{a}> 10^{9} {\rm GeV}$ we have
to conclude that the sphaleron induced damping dominates over the 
damping produced by the expansion of the Universe (i.e. $ \sim T^2/ M_{P}$).
The axion time evolution will then be (approximately) suppressed as
$\dot{\psi} \propto e^{- \gamma t} \sim e^{-\gamma \tau^2}$ in a radiation 
dominated Universe (recall that $t\sim \tau^2$). Therefore, we have to 
accept that in Eq. (\ref{decoupled}) the polarization mixing will be
 exponentially suppressed as time goes by. The sphaleron induced 
damping does not exclude other possible consequences associated with the
axions at the EWPT like the possible 
existence of strong CP violating effects \cite{kst}

Notice that this argument does to not apply to temperatures $T\ll T_{c}$ when  
the plasma has already cooled sufficiently for $\gamma$ to have turned off. 
Therefore, for our purposes it seems certainly theoretically more
plausible to consider the possible occurrence of coherent axion 
oscillations at temperatures of the oder of $1~{\rm GeV}$ when the electroweak 
symmetry is broken.

In can be, in principle, dangerous to have production of parity non invariant
 configurations of the ordinary magnetic field right before the onset of BBN. 
Suppose that the axion oscillations start at $T\sim 1 $ GeV with a typical mass
of $10^{-19}$ GeV. Therefore we have  to consider the ordinary axion  
oscillations coupled to the evolution of the ordinary vector potential at 
finite conductivity
\begin{eqnarray}
&&\ddot{\psi} + 3 H\dot{\psi} + m^2 \psi =0,
\nonumber\\
&&\sigma A_{\pm}' + \biggl[ k^2 \mp 
k\frac{\alpha_{{\rm em}}}{2 \pi} 
\frac{\psi'}{F_a}\biggr] A_{\pm} =0.
\label{axeq}
\end{eqnarray}
At high temperatures $T\gg \Lambda_{QCD}$ the axion is massless, but,
 at lower temperatures it develops a mass due to QCD instanton effects. 
The effects of instantons is highly suppressed at high temperatures, however,
the temperature dependence of the axion mass is approximately given by
\cite{gross}
\begin{equation}
m(T)\sim 0.1 \biggl(\frac{\Lambda_{QCD}}{T}\biggr)^{3.7} m_0,
\end{equation}
where $m_0$ is the axion mass at zero temperature.

Axion oscillations between a temperature of the order of $1$ GeV and 
$T\sim 10$ MeV (i. e. around the nucleosynthesis epoch) can generate 
electromagnetic helicity. Suppose, for sake of simplicity that the axion 
oscillations will start with a typical mass f the order of $10^{-19}$ GeV. 
Then, according to Eq. (\ref{axeq}), we have that the maximal amplified 
frequency and the diffusivity frequencies are of the order of
\begin{equation}
\omega_{{\rm max}} \sim \frac{\alpha_{{\rm em}}}{4 \pi} 
\frac{\Delta\psi}{F_a} \frac{T^2}{M_{0}}, 
~~~~\omega_{\sigma} \sim 
\sqrt{\frac{\sqrt{N_{eff}}}{\alpha_{{\rm em}}}}\sqrt{\frac{T}{M_{P}}}T.
\end{equation}
where $T$ is of the order of $1$ GeV. We can solve 
Eq. (\ref{axeq}) similarly to what we did in hypermagnetic case. In the case 
 $\omega_{\max} <\omega_{\sigma}$ the magnetic helicity produced 
via axion oscillations will be of the order of  
\begin{equation}
\frac{ \langle\vec{H} \cdot \vec{\nabla}\times\vec{H}\rangle }{\sigma\rho_c}
\simeq \frac{\alpha_{\rm em}}{4\pi} \frac{\Delta\psi}{F_{a}}\frac{T_a}{M_0}
\sim 10^{-22}
\label{al}
\end{equation}
(notice that $\sigma_0 \sim \alpha_{\rm em}^{-1}$ and $T_{a} \sim 1$ GeV).
Eq. (\ref{al}) 
provides then  a rather small parity breaking and certainly irrelevant for the 
dynamo mechanism.

\renewcommand{\theequation}{9.\arabic{equation}}
\setcounter{equation}{0}
\section{Concluding Remarks}

In this paper we investigated the production of hypermagnetic knots.
We found that HK can be efficiently produced provided some topologically
trivial configuration of the hypermagnetic field is already present for 
scales larger than the hypermagnetic diffusivity scale. The BAU 
can be successfully reproduced, but not in the MSM. The reason 
is twofold. On one hand the phase transition is too weak (even taking 
into account the possible presence of the hypermagnetic background). On the 
other hand the (perturbative) rate of the slowest processes in the plasma 
is too small. In the MSSM the BAU generation seems more probable and 
our mechanism can be viable, provided the typical scale of the HK 
exceeds the hypermagnetic diffusivity scale prior to the onset 
of the EWPT. Inspired by the 
axial Higgs we gave examples of specific 
scenarios where this mechanism can be implemented. We also 
discussed the possible production of magnetic helicity by  ordinary 
QCD axions: we found that its value is too small to be phenomenologically 
relevant.

\section*{Acknowledgments}
I am deeply indebted to M. Shaposhnikov for important discussions 
and valuable collaboration which motivated this investigation. 
I would also like to acknowledge interesting conversations with 
A. Vilenkin. Discussions with K. Jedamzik, H. Kurki-Suonio, 
J. Rehm and E. Sihvola were also appreciated. I would like 
to thank the Theory division of CERN and the Institute of Theoretical 
Physics of the Lausanne University where part of the work has been 
completed.

\newpage

\begin{appendix}

\renewcommand{\theequation}{B.\arabic{equation}}
\setcounter{equation}{0}
\section{Examples of Hypermagnetic Knots from Topologically non Trivial 
Configurations of Magnetic Fields}

\begin{figure}
\centerline{\epsfxsize = 12 cm  \epsffile{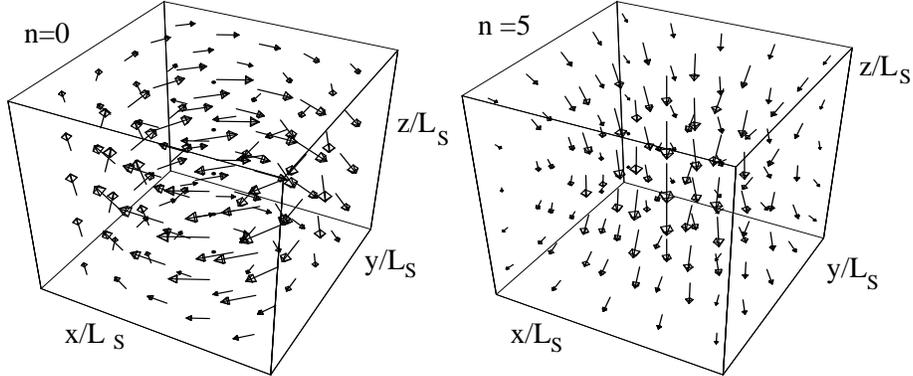}}
\caption[a]{ We plot the hypermagnetic knot configuration
 in Cartesian components 
[Eqs. (\ref{cart})] for the cases $n=0$ (zero helicity)
and $n= 5$. The
direction of the arrow at each point represents the tangent to the
flux lines of the magnetic field, whereas the length of the vector 
 is proportional to the field intensity. We notice that for large $n$ the field is
concentrated, in practice,  in the core of the
knot (i.e. $r/L_{s} <1$).}
\label{FIG1}
\end{figure}

In Section II we presented some topologically non trivial 
configurations of the hypercharge field  \cite{hyp,m1} whose 
main feature is to have a Chern-Simons number density 
which is time dependent and uniformly distributed in space.
It is also possible to construct hypermagnetic knot configurations 
with finite energy and helicity which are localized in space and within 
typical distance scale  $L_{s}$. 
Let us consider in fact the following configuration
in spherical coordinates \cite{ma,ran}
\begin{eqnarray}
{\cal Y}_{r}({\cal R},\theta) &=& - \frac{2 B_0}{ \pi L_{s}}
\frac{\cos{\theta} }{\bigl[{\cal R}^2 +1\bigr]^2},
\nonumber\\
{\cal Y}_{\theta}({\cal R},\theta) &=& \frac{2 B_0}
{ \pi L_{s}} \frac{ \sin{\theta}}{\bigl[ {\cal
R}^2 + 1\bigr]^2},
\nonumber\\
{\cal Y}_{\phi}({\cal R},\theta) &=& - \frac{ 2 B_0}{ \pi L_{s}} \frac{ n
{\cal
R}\sin{\theta}}{\bigl[{\cal R}^2 + 1\bigr]^2},
\label{conf2}
\end{eqnarray}
where ${\cal R}= r/L_{s}$ is the rescaled radius and $B_{0}$ is some 
dimensionless amplitude and $n$ is just an integer number 
whose physical interpretation will become clear in a moment. 
The hypermagnetic field can be easily computed 
from the previous expression and it is 
\begin{eqnarray}
&&{\cal H}_{r}({\cal R},\theta) = - \frac{4 B_{0}}{\pi~ L_{s}^2}\frac{n
\cos{\theta}}{\bigl[  {\cal
R}^2 + 1\bigr]^2},
\nonumber\\
&&{\cal H}_{\theta}({\cal R}, \theta) = - \frac{4 B_{0}}{\pi~
L_{s}^2}\frac{{\cal R}^2 -1}{\bigl[
{\cal R}^2 + 1\bigr]^3}n \sin{\theta},
\nonumber\\
&&{\cal H}_{\phi}({\cal R}, \theta) = 
- \frac{8 B_0}{ \pi~ L_{s}^2}\frac{ {\cal R}
\sin{\theta}}{\bigl[
{\cal R}^2 + 1\bigr]^3}.
\label{knot}
\end{eqnarray}
The poloidal and toroidal components of $\vec{{\cal H}}$ can be usefully 
expressed as $\vec{{\cal H}}_{p} = 
{\cal H}_{r} \vec{e}_{r} + {\cal H}_{\theta} \vec{e}_{\theta} $ 
and $\vec{\cal H}_{t}= {\cal H}_{\phi} \vec{e}_{\phi}$.
The Chern-Simons number is finite and it is given by 
\begin{equation}
N_{CS} =\frac{g'^2}{32\pi^2}
\int_{V} \vec{{\cal Y}} \cdot \vec{{\cal H}} d^3 x=
\frac{g'^2}{32\pi^2} \int_{0}^{\infty}
\frac{ 8 n
B^2_0}{\pi^2} \frac{ {\cal R}^2 d {\cal R}}{\bigr[ {\cal R}^2 +
1\bigl]^4} = \frac{g'^2 n B^2_0}{32 \pi^2}.
\label{CS}
\end{equation}
In Eq. (\ref{CS}) the integration is perfectly convergent over the whole
space and, since the field is $\vec{{\cal H}}_{Y}=0$ in any part of $\partial V$
at infinity, this quantity is also gauge invariant. Thus, we can see that 
the integer index $n$ labels the number of knots of the configuration.
We can also compute the total helicity of the configuration namely
\begin{equation}
\int_{V} \vec{{\cal H}}_{Y}
 \cdot \vec{\nabla} \times \vec{{\cal H}}_{Y} d^3 x=
\frac{256~B^2_0~n}{\pi L^2 } \int_{0}^{\infty} \frac{ {\cal R}^2 d
{\cal R}}{(1 + {\cal R}^2)^5} = \frac{5 B^2_0 n}{L_s^2}.
\label{helic}
\end{equation}
There is an important difference between Eqs. (\ref{CS}) and (\ref{helic}). In 
Eq. (\ref{CS}) the integrand is {\em not} gauge invariant (but the integrated 
Chern-Simons number density is gauge-invariant), on the other hand in Eq. (\ref{helic})
also the integrand is  gauge-invariant.
We can compute also the total energy of the field
\begin{equation}
E = \frac{1}{2}\int_{V} d^3 x |\vec{{\cal H}}_{Y}|^2 = \frac{B^2_0}{2~L_{s}}
(n^2 + 1).
\end{equation}
and we discover that it is proportional to $n^2$.
 This means that one way of increasing the total energy of
 the field is to increase the number of knots and twists in the flux lines.
We can also have some real space pictures of the core of the knot
 (i.e. ${\cal R} = r/L_{s}<1$). In Fig. \ref{FIG1} we plot the cartesian 
components of the hypermagnetic field 
\begin{eqnarray}
{\cal H}_{x}(\vec{{\cal X}}) &=& \frac{4 B_0}{\pi L_{s}^2}\frac{ 2 {\cal
 Y} - 2~ n~
{\cal X}{\cal Z}}{\bigl[1 + {\cal X}^2 + {\cal Y}^2 + {\cal
Z}^2\bigr]^3},
\nonumber\\
{\cal H}_{y}(\vec{{\cal X}}) &=& -
\frac{4 B_0}{\pi L_{s}^2}\frac{ 2 {\cal X} + 2~n~
{\cal Y} {\cal Z}}{\bigl[1 + {\cal X}^2 + {\cal Y}^2 + {\cal
Z}^2\bigr]^3},
\nonumber\\
{\cal H}_{z}(\vec{{\cal X}}) &=&
  \frac{4 B_0}{\pi L_{s}^2} \frac{ n \bigl[ {\cal
X}^2 + {\cal Y}^2 - {\cal Z}^2 -1\bigr]}{\bigl[1 + {\cal X}^2 + {\cal
Y}^2 + {\cal Z}^2\bigr]^3},
\label{cart}
\end{eqnarray}
in the case of $n=0$ and $n = 5$ in terms of the rescaled coordinates 
${\cal X} = x/L_{s}.~{\cal Y} = y/L_{s},~{\cal Z}=z/L_{s}$.

\renewcommand{\theequation}{B.\arabic{equation}}
\setcounter{equation}{0}
\section{Expectation Value of the Hypermagnetic Helicity}

The aim of this Appendix is to compute the expectation value of the 
hypermagnetic helicity which represents a fair gauge-invariant estimate
of parity breaking induced by the presence of dynamical psudoscalars in 
the hypermagnetic background.

Let us expand the hypercharge field operator according to Eq. 
(\ref{decomposition3}) but with slightly different notations which 
turn out to be useful for the calculation at hand:
\begin{equation}
\vec{Y}_{out}(\vec{x},\tau) = \frac{1}{\sqrt{2}} \sum_{\beta}\int \frac{d^3 k}
{\sqrt{(2\pi)^3}}\biggl(\hat{Y}_{\beta,out}(k,\tau)\vec{\epsilon}_{\beta} 
e^{i \vec{k}\cdot \vec{x}} +
\hat{Y}^{\dagger}_{\beta,out}(k,\tau)\vec{\epsilon}^{\ast}_{\beta} 
e^{-i \vec{k}\cdot \vec{x}}  \biggr)
\label{decomposition4}
\end{equation}
(where, as usual, $\beta= +,-$).

Therefore the expectation value of the hypermagnetic helicity will be
\begin{eqnarray}
\langle 0| \vec{H}_{Y}\cdot\vec{\nabla} \times \vec{H}_{Y}|
0 \rangle &=& 
\frac{1}{2} \int \frac{d^{3}k}{\sqrt{(2 \pi)^3}} 
\int \frac{d^{3}p}{\sqrt{(2 \pi)^3}} k p^2\biggl[ \langle\hat{Y}_{+}(k,\tau) 
\hat{Y}^{\dagger}_{+}(p, \tau) \rangle e^{ i (\vec{k} - \vec{p})\cdot\vec{x}}
- \langle\hat{Y}_{-}(k,\tau) 
\hat{Y}^{\dagger}_{-}(p, \tau) \rangle e^{ i (\vec{k} - \vec{p})\cdot\vec{x}}
\nonumber\\
&+&\langle\hat{Y}^{\dagger}_{+}(k,\tau) 
\hat{Y}_{+}(p, \tau) \rangle e^{ -i (\vec{k} - \vec{p})\cdot\vec{x}}
-\langle\hat{Y}^{\dagger}_{-}(k,\tau) 
\hat{Y}_{-}(p, \tau) \rangle e^{ -i (\vec{k} - \vec{p})\cdot\vec{x}}\biggr],
\end{eqnarray}
(we dropped the ``out'' subscript for the mode functions).
The various correlators appearing in the previous expression can be evaluated,
after some algebra, using the explicit form of the Bogoliubov transformation 
given in section II:
\begin{eqnarray}
&&\langle 0| \hat{Y}_{+}(k,\tau) 
\hat{Y}^{\dagger}_{+}(p, \tau) \rangle| 0\rangle 
\nonumber\\
&=&
\biggl[ \biggl(|c_{+}(k)|^2 + |c_{-}(k)|^2 \biggr)  f_{in}(\tau) 
f_{in}^{\ast}(\tau)  + c_{+}^{\ast}c_{-}(k){f_{in}^{\ast}(\tau)}^2 
+  c_{+}(k) c_{-}^{\ast}(k) f_{in}(\tau)^2\biggr] \delta^{(3)}(\vec{k} 
- \vec{p}),
\nonumber\\
&&\langle 0| \hat{Y}_{-}(k,\tau) 
\hat{Y}^{\dagger}_{-}(p, \tau) \rangle| 0\rangle 
\nonumber\\
&=&
\biggl[ \biggl(|\tilde{c}_{+}(k)|^2 + |\tilde{c}_{-}(k)|^2 \biggr)  
F_{in}(\tau) 
F_{in}^{\ast}(\tau)  
+ \tilde{c}_{+}^{\ast}\tilde{c}_{-}(k){F_{in}^{\ast}(\tau)}^2 
+  \tilde{c}_{+}(k) \tilde{c}_{-}^{\ast}(k) F_{in}(\tau)^2\biggr] 
\delta^{(3)}(\vec{k} - \vec{p}),
\nonumber\\
&&\langle 0| \hat{Y}_{+}^{\dagger}(k,\tau) 
\hat{Y}_{+}(p, \tau) \rangle| 0\rangle = \biggl(
\langle 0| \hat{Y}_{+}(k,\tau) 
\hat{Y}^{\dagger}_{+}(p, \tau) \rangle| 0
\rangle\biggr)^{\ast},
\nonumber\\
&&\langle 0| \hat{Y}^{\dagger}_{-}(k,\tau) 
\hat{Y}_{-}(p, \tau) \rangle| 0\rangle = 
\biggl( \langle 0| \hat{Y}_{-}(k,\tau) 
\hat{Y}^{\dagger}_{-}(p, \tau) \rangle| 0\rangle
\biggr)^{\ast}.
\end{eqnarray}
From these last expressions we get Eq. (\ref{expectation}).
With the same technique other expectation values of 
 parity non-invariant operators can be computed like
\begin{eqnarray}
\langle 0| \vec{H}_{Y}\cdot\vec{E}_{Y}|0\rangle &=&
 \frac{1}{2}\int\frac{d^{3} }{(2\pi)^3} k \biggl[ \biggl(
|c_{+}(k)|^2 +|c_{-}(k)|^2 \biggl) \biggl( f_{in}(\tau){f_{in}(\tau)^{\ast}}'+ 
f_{in}'(\tau){f_{in}(\tau)^{\ast}}\biggr) 
\nonumber\\
&&+ 2 c_{+}(k) c_{-}(k)^{\ast} f_{in}(\tau)' f_{in}(\tau) 
+2 c_{-}(k)c_{+}(k)^{\ast} {f_{in}(\tau)^{\ast}}' f_{in}(\tau)^{\ast}
\nonumber\\
&&- \biggl(
|\tilde{c}_{+}(k)|^2 +|\tilde{c}_{-}(k)|^2 \biggl) \biggl( 
F_{in}(\tau){F_{in}(\tau)^{\ast}}'+ F_{in}'(\tau){F_{in}(\tau)^{\ast}}\biggr) 
\nonumber\\
&&- 2 \tilde{c}_{+}(k) \tilde{c}_{-}(k)^{\ast} F_{in}(\tau)' F_{in}(\tau)
-2  \tilde{c}_{-}(k) \tilde{c}_{+}(k)^{\ast} {F_{in}(\tau)^{\ast}}' 
F_{in}(\tau)^{\ast} \biggr].
\end{eqnarray}

Notice that this second gauge-invariant operator can be also used in order to 
measure the amount of parity breaking. However, in the plasma it seems that 
the first operator computed in this Appendix is more suitable. Indeed, 
at finite conductivity, inhomogeneous hypermagnetic fields lead to the 
presence of induced (Ohm) hyperelectric fields of the order of
 $\vec{E}_{Y} \sim
\sigma^{-1} \vec{\nabla} \times \vec{H}_{Y}$.
\end{appendix}

\end{document}